\documentclass[sigconf]{acmart}
\AtBeginDocument{%
	\providecommand\BibTeX{{%
			\normalfont B\kern-0.5em{\scshape i\kern-0.25em b}\kern-0.8em\TeX}}}
\copyrightyear{2020}
\acmYear{2020}
\setcopyright{acmcopyright}
\acmConference[MM '20]{Proceedings of the 28th ACM International Conference on Multimedia}{October 12--16, 2020}{Seattle, WA, USA}
\acmBooktitle{Proceedings of the 28th ACM International Conference on Multimedia (MM '20), October 12--16, 2020, Seattle, WA, USA}
\acmPrice{15.00}
\acmDOI{10.1145/3394171.3413564}
\acmISBN{978-1-4503-7988-5/20/10}

\usepackage{multirow}
\begin{document}
	\fancyhead{}
	\title{Attention Cube Network for Image Restoration}
\author{Yucheng Hang}
\affiliation{%
	\institution{Shenzhen International Graduate School, Tsinghua University}
	\city{Shenzhen}
	\country{China}}
\email{ychang20@163.com}

\author{Qingmin Liao}
\affiliation{%
	\institution{Shenzhen International Graduate School $\&$ Department of Electronic Engineering, Tsinghua University}
	\city{Shenzhen}
	\country{China}}
\email{liaoqm@tsinghua.edu.cn}

\author{Wenming Yang}
\authornote{Corresponding author.}
\affiliation{%
	\institution{Shenzhen International Graduate School $\&$ Department of Electronic Engineering, Tsinghua University}
	\city{Shenzhen}
	\country{China}}
\email{yang.wenming@sz.tsinghua.edu.cn}

\author{Yupeng Chen}
\affiliation{%
	\institution{Peng Cheng Laboratory}
	\city{Shenzhen}
	\country{China}}
\email{chenyp01@pcl.ac.cn}

\author{Jie Zhou}
\affiliation{%
	\institution{Department of Automation, Tsinghua University}
	\city{Beijing}
	\country{China}}
\email{jzhou@tsinghua.edu.cn}
	\begin{abstract}
		Recently, deep convolutional neural network (CNN) have been widely used in image restoration and obtained great success. However, most of existing methods are limited to local receptive field and equal treatment of different types of information. Besides, existing methods always use a multi-supervised method to aggregate different feature maps, which can not effectively aggregate hierarchical feature information. To address these issues, we propose an attention cube network (A-CubeNet) for image restoration for more powerful feature expression and feature correlation learning. Specifically, we design a novel attention mechanism from three dimensions, namely spatial dimension, channel-wise dimension and hierarchical dimension. The adaptive spatial attention branch (ASAB) and the adaptive channel attention branch (ACAB) constitute the adaptive dual attention module (ADAM), which can capture the long-range spatial and channel-wise contextual information to expand the receptive field and distinguish different types of information for more effective feature representations. Furthermore, the adaptive hierarchical attention module (AHAM) can capture the long-range hierarchical contextual information to flexibly aggregate different feature maps by weights depending on the global context. The ADAM and AHAM cooperate to form an "attention in attention" structure, which means AHAM's inputs are enhanced by ASAB and ACAB. Experiments demonstrate the superiority of our method over state-of-the-art image restoration methods in both quantitative comparison and visual analysis.
	\end{abstract}
	\begin{teaserfigure}
		\includegraphics[width=\textwidth]{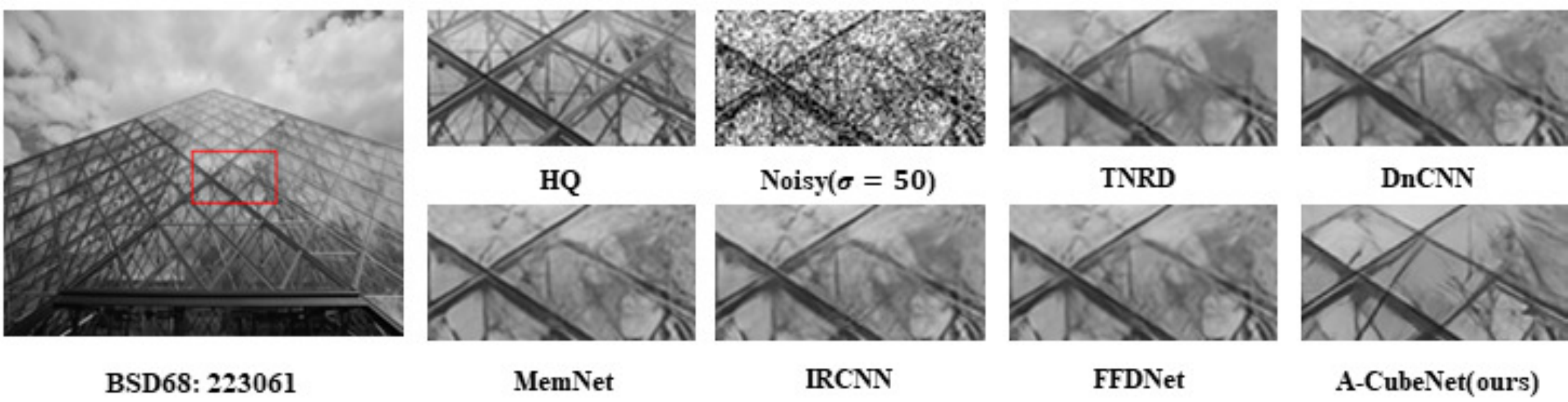}
		\caption{Gray image denoising results with noise level 50 on “BSD: 223061” from BSD68. Our method obtains better visual quality and recovers more textural details compared with other state-of-the-art methods.}
		\Description{Gray image denoising results with noise level 50 on “BSD: 223061” from BSD68. Our method obtains better visual quality and recovers more textural details compared with other state-of-the-art methods.}
		\label{fig:teaser}
	\end{teaserfigure}
	\maketitle
	\section{Introduction}
	Image restoration is a classic computer vision task that aims to recover high-quality images from low-quality images corrupted by various kinds of degradations. Due to the irreversible nature of the image degradation process, it is an ill-posed problem. It can be categorized into different tasks such as image super-resolution, image denoising, JPEG image deblocking, etc. \par
	Recently, methods based on deep convolutional neural network (CNN) have been widely used in image restoration due to their strong nonlinear representational power. \cite{dong2015image,kim2016accurate,kim2016deeply,haris2018deep,zhang2018residual,zhang2018image,dai2019second,hui2019lightweight} focused on designing a deeper, wider or lighter network structure, aiming at improving the performance of image super-resolution. Dong et al.\cite{dong2015compression} proposed ARCNN with several stacked convolutional layers for JPEG image deblocking. By taking a tunable noise level map as input, FFDNet can deal with noise on different levels \cite{zhang2018ffdnet}. Guo et al.\cite{guo2019toward} proposed CBDNet for blind denoising of real images. However, these methods are for specific image restoration tasks. Different from them, \cite{liu2018non,liu2018multi,zhang2017beyond,zhang2017learning,zhang2019residual} developed a couple of valuable methods that can be generalized to different image restoration tasks. \par
	Although above CNN-based image restoration methods have achieved gratifying results, they still have some problems: (1) The receptive fields of these methods are relatively small. Most of these methods extract features in a local way through convolution operations, which cannot capture the long-range dependencies between pixels in the whole image. A larger receptive field can make better use of training images and more contextual information, which is very helpful for image restoration, especially when the images suffer from heavy corruptions. (2) Most of these methods treat all types of information (e.g., low and high frequency information) equally, which may result in over-smoothed reconstructed images and fail to recover some textural details. In other words, the power of discrimination of these methods is limited. (3) In response to problem one and two, although there have been recent methods that use attention mechanism (e.g., SENet \cite{hu2018squeeze} and non-local neural network \cite{wang2018non}) for image super-resolution, most methods directly introduce the attention mechanism in high-level computer vision tasks and ignore the difference between high-level and low-level computer vision tasks \cite{zhang2018image,dai2019second}. (4) Most of these methods only use the feature map outputted from the last layer for image restoration. In fact, in deep convolutional neural networks, feature map information at different levels can complement each other. If only the feature map outputted from the last layer is used for image restoration, part of the information is bound to be lost. Although there are some methods that take this problem into account, they just cascade them together \cite{zhang2018residual} or use a  multi-supervised method \cite{kim2016deeply,tai2017memnet}, and the feature map information of each level is not fully and effectively used. \par
	In order to solve above problems, this paper proposes an attention cube network (A-CubeNet) for image restoration based on adaptive dual attention module (ADAM) and adaptive hierarchical attention module (AHAM). Specifically, this paper proposes an adaptive dual attention module for above mentioned problem one, two, and three. This module can capture the long-range dependencies between pixels and channels to expand the receptive field and distinguish different types of information for more effective feature representations, which is very helpful for the restoration of textural details. In addition, inspired by the non-local neural network, this paper designs an adaptive hierarchical attention module for above mentioned problem four. This module first performs squeeze operations on the spatial and channel-wise dimensions of each feature map for global context modeling. Then this module captures the long-range dependencies between different feature maps. Finally, this module fuses each feature map based on the long-range dependencies between different feature maps. Full experiments show that compared with other state-of-the-art methods, our A-CubeNet achieves the best results in all tasks. \par
	In summary, the main contributions of this paper are listed as follows: \par
	$\bullet$ We propose an adaptive dual attention module (ADAM), including an adaptive spatial attention branch (ASAB) and an adaptive channel attention branch (ACAB). ADAM can capture the long-range spatial and channel-wise contextual information to expand the receptive field and distinguish different types of information for more effective feature representations. Therefore our A-CubeNet can obtain high-quality image restoration results, as shown in Figure \ref{fig:teaser}. \par
	$\bullet$ Inspired by the non-local neural network, we design an adaptive hierarchical attention module (AHAM), which flexibly aggregates all output feature maps together by the hierarchical attention weights depending on global context. To the best of our knowledge, this is the first time to consider aggregating output feature maps in a hierarchical attention method with global context. \par
	$\bullet$ Through sufficient experiments, we prove that our A-CubeNet is powerful for various image restoration tasks. Compared with other state-of-the-art methods for image denoising, JPEG image deblocking and image super-resolution, our A-CubeNet obtains superior results in both quantitative metrics and visual quality.
	\section{related work}
	\subsection{Image Restoration}
	CNN-based image restoration methods cast image restoration as an image-to-image regression problem, and learn an end-to-end mapping from low-quality (LQ) to high-quality (HQ) directly. Dong et al. \cite{dong2014learning,dong2015compression} proposed SRCNN for image super resolution and ARCNN for JPEG image deblocking. Both of them achieved superior performance against previous methods that are not based on CNN. Then \cite{kim2016accurate,kim2016deeply,haris2018deep,zhang2018residual,zhang2018image,dai2019second,hu2019channel} focused on designing a deeper and wider network structure or considering feature correlations to improve the performance of image super-resolution. Zhang et al. \cite{zhang2017beyond} proposed DnCNN for image denoising and JPEG image deblocking by introducing residual learning to train deeper network. \cite{zhang2017learning} introduced the denoiser prior for fast image restoration. Tai et al. \cite{tai2017memnet} lately designed a persistent memory network for image restoration and achieved promising results. However, most methods neglect to consider feature correlations in spatial or channel-wise dimensions and can not make full use of hierarchical feature maps.
	\subsection{Attention Mechanism}
	Attention mechanism is inspired by the cognitive process of human \cite{mnih2014recurrent}. Human always focuses on more important information. Attention mechanism has been widely used in various computer vision tasks, such as image and video classification tasks \cite{hu2018squeeze,wang2018non}. Wang et al. \cite{wang2018non} proposed non-local neural network by incorporating non-local operations for spatial attention in video classification. \cite{hu2018squeeze} modelled channel-wise relationships to obtain significant performance gain for image classification. Woo et al. \cite{woo2018cbam} developed CBAM to model both channel-wise and spatial relationships. In all, these works mainly aimed to concentrate on more useful information in features. \par
	Recently, SENet was introduced to improve SR performance \cite{zhang2018image}. \cite{hu2019channel,kim2018ram,dai2019second} focused on introducing both channel attention and spatial attention to image super-resolution. Dai et al. \cite{dai2019second} designed a second-order attention mechanism based on SENet and introduced non-local neural network to further improve SR performance. Following the importance of self-similarity prior, \cite{zhang2019residual} adapted non-local operations into their network. However, most methods directly introduce the attention mechanism in high-level computer vision tasks and ignore the difference between high-level and low-level computer vision tasks. For example, non-local neural network is computationally intensive and is actually not suitable for image restoration.
	\section{method}
	\subsection{Framework}
\begin{figure*}[htbp]
		\centering
		\includegraphics[width=16cm]{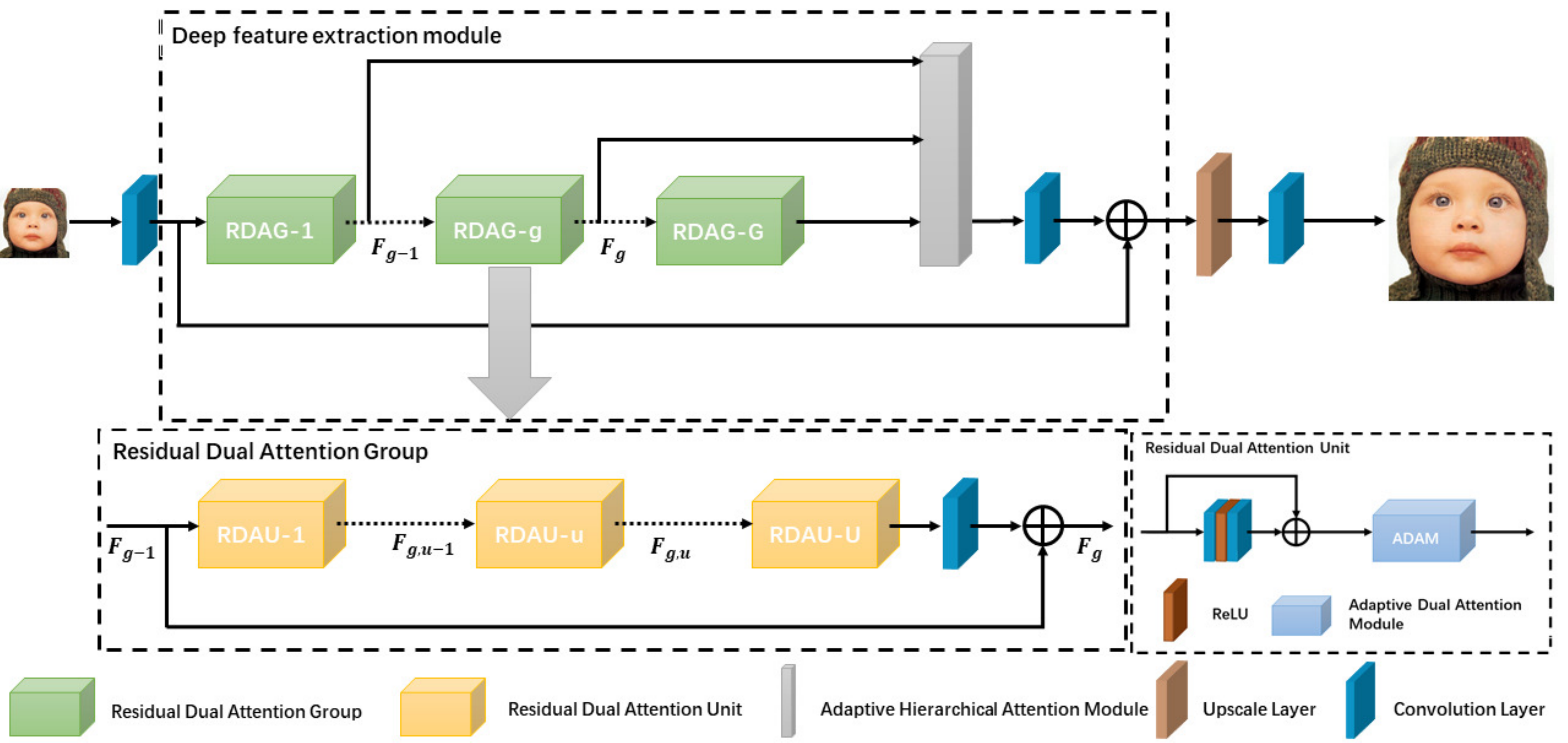}
		\caption{Framework of the proposed attention cube network (A-CubeNet)}
		\Description{Framework of the proposed attention cube network (A-CubeNet)}
		\label{fig:framework}
	\end{figure*}

	As shown in Figure \ref{fig:framework}, the proposed A-CubeNet mainly consists of three modules: the shallow feature extraction module, the deep feature extraction module stacked with G residual dual attention groups (RDAGs) and one adaptive hierarchical attention module (AHAM), and the construction module. Given $I_{L Q}$ and $I_{H Q}$ as the low-quality (e.g., noisy, low resolution, or compressed images) and high-quality images. We apply only one convolutional layer to extract the shallow feature $F_{0}$ from the low-quality input:
	\begin{equation}
	F_{0}=H_{S F}\left(I_{L Q}\right)\text{, }
	\end{equation}
	where $H_{S F}(\cdot)$ represents the shallow feature extraction module. Then the extracted shallow feature $F_{0}$ is used for RDAGs and AHAM based deep feature extraction, which thus produces the deep feature as:
	\begin{equation}
	F_{D F}=H_{D F}\left(F_{0}\right)=F_{0}+W_{L S C} H_{A H A M}\left(F_{1}, \cdots, F_{g}, \cdots, F_{G}\right)\text{, }
	\end{equation}
	where $H_{D F}(\cdot)$, $W_{L S C}$ and $H_{A H A M}(\cdot)$ represent the deep feature extraction module, the convolutional layer at the end of the deep feature extraction module and the adaptive hierarchical attention module (AHAM), respectively. To address the image super-resolution task, we add an extra upscale layer before the last convolutional layer. Specifically, we utilize sub-pixel convolutional operation (convolution + pixel shuffle) \cite{shi2016real} to upscale feature maps:
	\begin{equation}
	F_{U P}=H_{U P}\left(F_{D F}\right)\text{, }
	\end{equation}
	where $H_{U P}(\cdot)$ represents the upscale module. Then we use a convolutional layer to get the reconstructed image:
	\begin{equation}
	I_{R E C}=H_{R E C}\left(F_{D F}\right) \text { or } I_{R E C}=H_{R E C}\left(F_{U P}\right)\text{, }
	\end{equation}
	where $H_{R E C}(\cdot)$ represents the reconstruction module. The overall reconstruction process can be expressed as:
	\begin{equation}
	I_{R E C}=H_{A-CubeNet}\left(I_{L Q}\right)\text{, }
	\end{equation}
	where $H_{A-CubeNet}(\cdot)$ represents the function of our A-CubeNet. \par
	Then A-CubeNet is optimized with a certain loss function. Some loss functions have been widely adopted, such as L2 \cite{zhang2017beyond,zhang2017learning,tai2017memnet,dong2015compression}, L1 \cite{lim2017enhanced,zhang2018residual,dai2019second}, perceptual and adversarial losses \cite{ledig2017photo}. To verify the effectiveness of our A-CubeNet, we adopt the same loss functions as previous works (e.g., L1 loss function for image super-resolution, L2 loss function for image denoising and JPEG image deblocking). Given a training set $\left\{I_{L Q}^{i}, I_{H Q}^{i}\right\}_{i=1}^{N}$ with N LQ images and their HQ counterparts. The goal of training A-CubeNet is to optimize the loss function:
	\begin{equation}
	\begin{aligned}
	&L_{1}(\theta)=\frac{1}{N} \sum_{i=1}^{N}\left\|H_{A-CubeNet}\left(I_{L Q}^{i}\right)-I_{H Q}^{i}\right\|_{1}\text{, }\\
	&L_{2}(\theta)=\frac{1}{N} \sum_{i=1}^{N}\left\|H_{A-CubeNet}\left(I_{L Q}^{i}\right)-I_{H Q}^{i}\right\|_{2}\text{, }
	\end{aligned}
	\end{equation}
	where $\theta$ indicates the updateable parameters of our A-CubeNet.
	\subsection{Residual Dual Attention Group (RDAG)}
	As shown in Figure \ref{fig:framework}, the deep feature extraction module stacked with several RDAGs, while an RDAG consists of two parts: U stacked RDAUs and one convolutional layer. The g-th RDAG can be expressed as:
	\begin{equation}
	F_{g}=F_{g-1}+W_{S S C} H_{g, u}\left(H_{g, u-1}\left(\cdots H_{g, 1}\left(F_{g-1}\right) \cdots\right)\right)\text{, }
	\end{equation}
	where $F_{g-1}$ and $F_{g}$ represent the input and output of the g-th RDAG. $W_{S S C}$ and $H_{g, u}(\cdot)$ represent the convolutional layer at the end of the g-th RDAG and the u-th RDAU of the g-th RDAG, respectively. Each RDAU consists of one residual block and one adaptive dual attention module (ADAM). Then we give more details to ADAM and AHAM.
	\subsection{Adaptive Dual Attention Module (ADAM)}
	\begin{figure}[htbp]
		\centering
		\includegraphics[width=\linewidth]{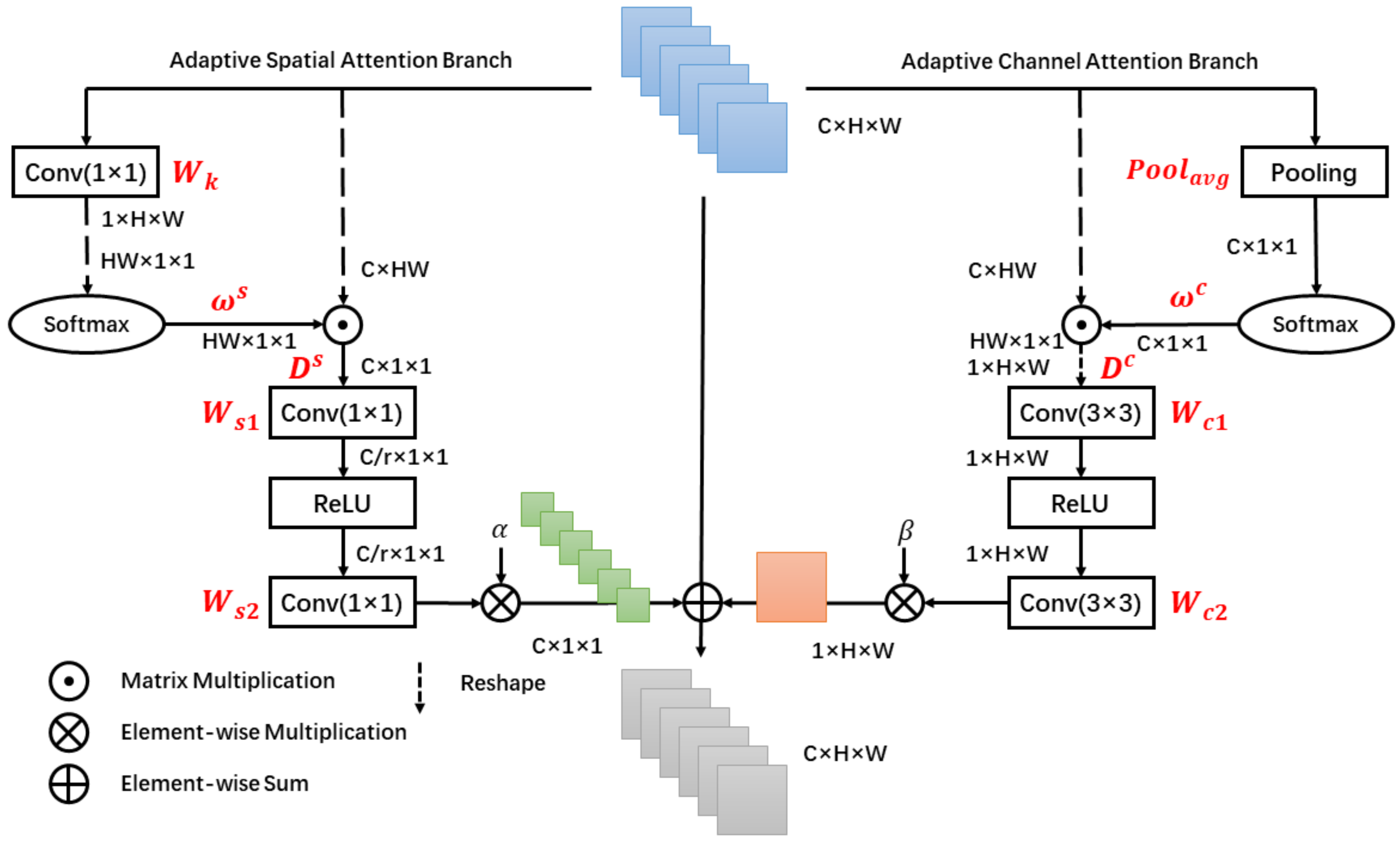}
		\caption{Adaptive dual attention module (ADAM)}
		\Description{Adaptive dual attention module}
		\label{fig:attention}
	\end{figure}
	As shown in Figure \ref{fig:attention}, our adaptive dual attention module (ADAM) consists of two branches: adaptive spatial attention branch (ASAB) and adaptive channel attention branch (ACAB). These two branches cooperate to draw global features rather than local features and endow the network the ability to treat different types of information differently. Thus our A-CubeNet obtains better feature representations for high-quality image restoration. Take the adaptive spatial attention branch in Figure \ref{fig:attention} as an example. First, we apply a convolutional layer to squeeze channel-wise features and apply softmax function to obtain the attention weights. Second, we perform a matrix multiplication between the attention weights and the original features to obtain long-range spatial contextual information. Third, we perform feature transform and feature fusion with adaptive weight. Similarly, we obtain channel-wise long-range contextual information by the adaptive channel attention branch. In general, each branch has three steps: (1) Squeeze features; (2) Extract long-range contextual information; (3) Perform feature fusion with adaptive weight. Our ADAM can be formulated as:
	\begin{equation}
	O u t_{A D A M}=X+O u t_{A S A B}+O u t_{A C A B}\text{, }
	\end{equation}
	where X , $O u t_{A S A B}$, $O u t_{A C A B}$ and $O u t_{A D A M}$ represent the input feature map, the output of ASAB, the output of ACAB and the output feature map of ADAM, respectively.
	\subsubsection{Adaptive spatial attention branch (ASAB)}
	As shown in Figure \ref{fig:attention}, given an input feature map $X=\left\{X_{i}\right\}_{i=1}^{N}$, $X \in \mathbb{R}^{C \times H \times W}$, where $N=H \times W$ is the number of pixels. First, we apply a convolutional layer $W_{k}$ to squeeze channel-wise features: $A^{s}=W_{k} X$, $A^{s} \in \mathbb{R}^{1 \times H \times W}$. Then we reshape it to $\mathbb{R}^{H W \times 1 \times 1}$ and apply a softmax fuction to obtain the spatial attention weights $\omega^{s} \in \mathbb{R}^{H W \times 1 \times 1}$: 
	\begin{equation}
	\omega^{s}_{j}=\frac{\exp \left(A^{s}_{j}\right)}{\sum_{m=1}^{N} \exp \left(A^{s}_{m}\right)}\text{, }
	\end{equation}
	Then we reshape $X \in \mathbb{R}^{C \times H \times W}$ to obtain $B^{s} \in \mathbb{R}^{C \times H W}$. After that we perform long-range spatial contextual information modeling, which groups the features of all pixels together with the spatial attention weights to obtain the long-range spatial contextual features. Specifically, we perform a matrix multiplication between $B^{s}$ and the spatial attention weights $\omega^{s}$:
	\begin{equation}
	D^{s}=\sum_{j=1}^{N} \omega^{s}_{j} B^{s}_{j}\text{, }
	\end{equation}
	where $D^{s} \in \mathbb{R}^{C \times 1 \times 1}$ represents spatial long-range contextual features. Then we feed it into a bottleneck network (e.g., one $1 \times 1$ convolutional layer $W_{s1}$, one ReLU activation layer and one $1 \times 1$ convolutional layer $W_{s2}$) to perform feature transform. Finally, we multiply it by an adaptive learning weight $\alpha$:
	\begin{equation}
	\begin{aligned}
	O u t_{A S A B}&=\alpha W_{s2} \operatorname{ReLU}\left(W_{s1} \sum_{j=1}^{N} \omega^{s}_{j} B^{s}_{j}\right)\\
	&=\alpha W_{s2} \operatorname{ReLU}\left(W_{s1} \sum_{j=1}^{N} \frac{\exp \left(W_{k} X_{j}\right)}{\sum_{m=1}^{N} \exp \left(W_{k} X_{m}\right)} X_{j}\right)\text{. }
	\end{aligned}
	\label{equ:asab}
	\end{equation}
	where $O u t_{A S A B} \in \mathbb{R}^{C \times 1 \times 1}$ represents the output of ASAB. \par
	We emphasize three points: (1) $\alpha$ is initialized as 0 and gradually learns to assign larger weight. This adaptive learning weight helps our A-CubeNet fuse long-range spatial contextual features effectively. (2) We use a bottleneck network with the bottleneck ratio r instead of only one $1 \times 1$ convolutional layer because: (a) Compared with using only one $1 \times 1$ convolutional layer, the number of parameters reduces from $C^2$ to $2C^2/r$. (b) Just like SENet \cite{hu2018squeeze}, it can better fit the complex correlation between channels with more nonlinearity. However, noted that this correlation between channels is directed at long-range spatial contextual features instead of the input feature map. So it can only be regarded as a supplement to the spatial attention mechanism. (3) The Equation \ref{equ:asab} shows that $O u t_{A S A B}$ is a weighted sum of the features across all pixels. Therefore, it has a global receptive field and selectively aggregates context according to the spatial attention weights.
	\subsubsection{Adaptive channel attention branch (ACAB)}
	As shown in Figure \ref{fig:attention}, given an input feature map $X=\left\{X_{i}\right\}_{i=1}^{C}$, $X \in \mathbb{R}^{C \times H \times W}$, where C is the number of channels. First, we apply an average pooling layer ${Pool}_{\text {avg }}$ to squeeze spatial features: $A^{c}={Pool}_{\text {avg }} X$, $A^{c} \in \mathbb{R}^{C \times 1 \times 1}$. Then we apply a softmax fuction to obtain the channel-wise attention weights $\omega^{c} \in \mathbb{R}^{C \times 1 \times 1}$:
	\begin{equation}
	\omega^{c}_{j}=\frac{\exp \left(A^{c}_{j}\right)}{\sum_{m=1}^{C} \exp \left(A^{c}_{m}\right)}\text{, }
	\end{equation}
	Then we reshape $X \in \mathbb{R}^{C \times H \times W}$ to obtain $B^{c} \in \mathbb{R}^{C \times H W}$. After that we perform long-range channel-wise contextual information modeling, which groups the features of all channels together with the channel-wise attention weights to obtain long-range channel-wise contextual features. Specifically, we perform a matrix multiplication between $B^{c}$ and the channel-wise attention weights $\omega^{c}$:
	\begin{equation}
	D^{c}=\sum_{j=1}^{C} \omega^{c}_{j} B^{c}_{j}\text{, }
	\end{equation}
	where $D^{c} \in \mathbb{R}^{C \times 1 \times 1}$ represents channel-wise long-range contextual features. Then we feed it into a network (e.g., one $3 \times 3$ convolutional layer $W_{c1}$, one ReLU activation layer and one $3 \times 3$ convolutional layer $W_{c2}$) to perform feature transform. Finally, we multiply it by an adaptive learning weight $\beta$:
	\begin{equation}
	\begin{aligned}
	O u t_{A C A B}&=\beta W_{c2} \operatorname{ReLU}\left(W_{c1} \sum_{j=1}^{C} \omega^{c}_{j} B^{c}_{j}\right)\\
	&=\beta W_{c2} \operatorname{ReLU}\left(W_{c1} \sum_{j=1}^{C} \frac{\exp \left(\operatorname{Pool}_{\text {avg }} X_{j}\right)}{\sum_{m=1}^{C} \exp \left(\operatorname{Pool}_{\text {avg }} X_{m}\right)} X_{j}\right)\text{. }
	\end{aligned}
	\label{equ:acab}
	\end{equation}
	where $O u t_{A C A B} \in \mathbb{R}^{1 \times H \times W}$ represents the output of ACAB. \par
	We emphasize three points: (1) $\beta$ is initialized as 0 and gradually learns to assign larger weight. This adaptive learning weight helps our AGAM fuse long-range channel-wise contextual features effectively. (2) We use a network instead of only one $3 \times 3$ convolutional layer because it can better fit the complex spatial correlation with more nonlinearity, just like CSAR block \cite{hu2019channel}. However, noted that this spatial correlation is directed at long-range channel-wise contextual features instead of the input feature map. So it can only be regarded as a supplement to the channel attention mechanism. (3) The Equation \ref{equ:acab} shows that $O u t_{A C A B}$ is a weighted sum of the features across all channels. It makes our A-CubeNet focus on more informative features and improve the power of discrimination.
	\subsection{Adaptive Hierarchical Attention Module (AHAM)}
	\begin{figure}[htbp]
		\centering
		\includegraphics[width=\linewidth]{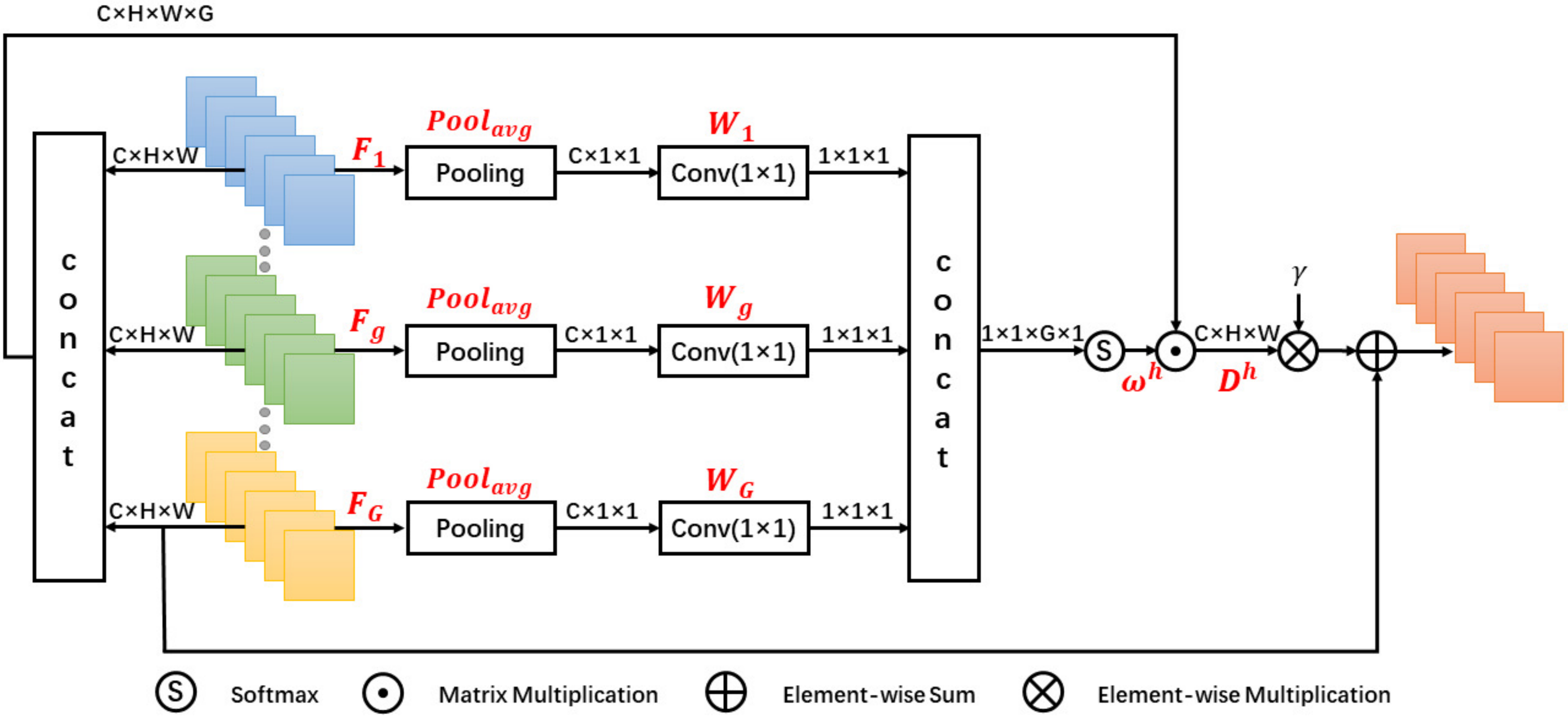}
		\caption{Adaptive hierarchical attention module (AHAM)}
		\Description{Adaptive hierarchical attention module}
		\label{fig:gcfm}
	\end{figure}
	As shown in Figure \ref{fig:gcfm}, given a set of each residual dual attention group's output feature map $F=\left\{F_{g}\right\}_{g=1}^{G}$, $F_{g} \in \mathbb{R}^{C \times H \times W}$, where G is the number of residual dual attention groups (RDAGs). First, we apply an average pooling layer ${Pool}_{\text {avg }}$ and a convolutional layer $W_{g}$ to squeeze global features of each feature map: $A^{h}_{g}=W_{g}{Pool}_{\text {avg }} F_{g}$, $A^{h}_{g} \in \mathbb{R}^{1 \times 1 \times 1}$. Then we cascade them to obtain $A^{h} \in \mathbb{R}^{1 \times 1 \times G \times 1}$. After that we apply a softmax function to obtain the hierarchical attention weights $\omega^{h} \in \mathbb{R}^{1 \times 1 \times G \times 1}$:
	\begin{equation}
	\omega^{h}_{j}=\frac{\exp \left(A^{h}_{j}\right)}{\sum_{m=1}^{G} \exp \left(A^{h}_{m}\right)}\text{, }
	\end{equation}
	Then we cascade each $F_{g}$ to obtain $B^{h} \in \mathbb{R}^{C \times H \times W \times G}$. After that we perform global contextual information modeling, which groups all RDAG’s output feature maps together with the self-adaptive weights to obtain global contextual features. Specifically, we perform a matrix multiplication between $B^{h}$ and the hierarchical attention weights $\omega^{h}$:
	\begin{equation}
	D^{h}=\sum_{j=1}^{G} \omega^{h}_{j} B^{h}_{j}\text{, }
	\end{equation}
	where $D^{h} \in \mathbb{R}^{C \times H \times W}$ represents global contextual features. Then, we multiply it by an adaptive learning weight $\gamma$. Finally, we perform an element-wise sum operation with the last RDAG’s output feature map $F_{G}$ to obtain the final output $O u t_{A H A M} \in \mathbb{R}^{C \times H \times W}$:
	\begin{equation}
	\begin{aligned}
	O u t_{A H A M}&=F_{G}+\gamma \sum_{j=1}^{G} \omega^{h}_{j} B^{h}_{j}\\
	&=F_{G}+\gamma \sum_{j=1}^{G} \frac{\exp \left(W_{j} \operatorname{Pool}_{\text {avg }} F_{j}\right)}{\sum_{m=1}^{G} \exp \left(W_{m} \operatorname{Pool}_{\text {avg }} F_{m}\right)} F_{j}\text{. }
	\end{aligned}
	\label{equ:gcfm}
	\end{equation} \par
	We emphasize three points: (1) $\gamma$ is initialized as 0 and gradually learns to assign larger weight. This adaptive learning weight helps our A-CubeNet fuse global contextual features effectively. (2) The Equation \ref{equ:gcfm} shows that $O u t_{A H A M}$ is a weighted sum of all RDAG’s output feature maps. The hierarchical attention weights depend on global context of all intermediate feature maps. (3) AHAM can learn how to combine different hierarchical feature maps that are most conducive to reconstruction.
	\section{experiments}
	\subsection{Datasets and Metrics}
	We apply our A-CubeNet to three classical image restoration tasks: image super-resolution, gray image denoising and JPEG image deblocking. DIV2K dataset \cite{agustsson2017ntire} is used to train all of our models. For image super-resolution, we follow the same settings as EDSR \cite{lim2017enhanced}. Set5 \cite{BMVC.26.135}, Set14 \cite{Zeyde2010On}, BSD100 \cite{Martin2002A}, Urban100 \cite{Huang2015Single} and Manga109 \cite{Yusuke2017Sketch} are adopted as the test datasets. For gray image denoising, we follow the same setting as IRCNN \cite{zhang2017learning}. BSD68 \cite{Martin2002A} and Kodak24 are used as the test datasets. For JPEG image deblocking, we follow the same setting as ARCNN \cite{dong2015compression}. LIVE1 \cite{Moorthy2009Visual} and Classic5 \cite{Foi2007Pointwise} are applied as the test datasets. For each task, We adopt the mean PSNR and/or SSIM to evaluate the results.
	\subsection{Implementation Details}
	Our A-CubeNet contains 4 RDAGs (G = 4) and each RDAG contains 4 RDAUs (U = 4). In each ASAB, we use $1 \times 1$ convolutional filter with the bottleneck ratio r = 16. In each ACAB, we set the size and number of filters as $3 \times 3$ and 1. In AHAM, we set the size and number of filters as $1 \times 1$ and 1. For other convolutional layers, the size and number of filters are set as $3 \times 3$ and 64.
	During training, data augmentation is performed on the training images, which are randomly rotated by $90^{\circ}$, $180^{\circ}$, $270^{\circ}$ and flipped horizontally. In each min-batch, we randomly cropped 16 patches with size 48×48 from the low-quality (LQ) images. Our model is trained by ADAM optimizer with $\beta_{1}=0.9$, $\beta_{2}=0.999$ and $\epsilon=10^{-8}$. The learning rate is initialized as $2 \times 10^{-4}$ and then decreases to half every $2 \times 10^{5}$ iterations of back-propagation. We use PyTorch framework to implement our A-CubeNet with a GTX 1080Ti GPU.
	\subsection{Ablation Study}
	\subsubsection{Adaptive dual attention module (ADAM)}
	We follow the same ablation study setting as RAM \cite{kim2018ram} to compare different attention mechanisms conveniently. Specifically, we set a baseline network with 16 residual blocks. Then we implement the attention mechanisms in each residual block. For fair comparison, all networks have the same hyperparameter settings and use the same training and testing process.
	We compare our ADAM with RAM \cite{kim2018ram}, RCAB \cite{zhang2018image}, CBAM \cite{woo2018cbam} and CSAR \cite{hu2019channel}. As shown in Table \ref{tab:ablotionAMAM}, our ADAM achieves the best PSNR results on all datasets (the average gain is 0.23dB) with only 10K additional parameters (which is an increase of 0.7\%). \par
	Furthermore, we conduct ablation study inside our ADAM. As shown in Table \ref{tab:ablation inside ADAM},  we set up ADAM-S, which only contains ASAB and the corresponding parameter $\alpha$. Similarly, we design ADAM-C, which only contains ACAB and the corresponding parameter $\beta$. Compared with the baseline network, ADAM-C and ADAM-S increase 0.08dB and 0.11dB, which verifies the effectiveness of ACAB and ASAB respectively. Then we propose ADAM-NW, which removes the adaptive learning weights $\alpha$ and $\beta$. Compared with ADAM, ADAM-NW decreases 0.09dB, which proves that the adaptive learning weights $\alpha$ and $\beta$ are quite effective. \par
	In summary, our ADAM obtains a significant improvement for image restoration with a little additional parameters. Therefore, it can be inserted into most CNN-based image restoration methods.
	\begin{table}[htbp]
		\centering
		\caption{Performance of our ADAM and other attention mechanisms for image super-resolution with scaling factor $\times$2. \textcolor[rgb]{ 1,  0,  0}{Red} and \textcolor[rgb]{ 0,  0,  1}{blue} colors indicate the best and second best performance, respectively.}
		\label{tab:ablotionAMAM}
		\resizebox{\linewidth}{1.5cm}{
			\begin{tabular}{ccccccc}
				\toprule
				\multicolumn{1}{c}{} & \multicolumn{1}{c}{Baseline} & \multicolumn{1}{c}{+ADAM} & \multicolumn{1}{c}{+RAM} & \multicolumn{1}{c}{+RCAB} & \multicolumn{1}{c}{+CBAM} & \multicolumn{1}{c}{+CSAR} \\
				\midrule
				Params. & \multicolumn{1}{c}{1370K} & \multicolumn{1}{c}{1380K} & \multicolumn{1}{c}{1389K} & \multicolumn{1}{c}{1379K} & \multicolumn{1}{c}{1381K} & \multicolumn{1}{c}{1646K} \\
				Set5  & 37.90 & \textcolor[rgb]{ 1,  0,  0}{38.10} & \textcolor[rgb]{ 0,  0,  1}{37.98} & 37.96 & 37.89 & 37.96 \\
				Set14 & 33.58 & \textcolor[rgb]{ 1,  0,  0}{33.72} & 33.57 & \textcolor[rgb]{ 0,  0,  1}{33.58} & 33.45 & 33.57 \\
				BSD100 & 32.17 & \textcolor[rgb]{ 1,  0,  0}{32.25} & \textcolor[rgb]{ 0,  0,  1}{32.17} & \textcolor[rgb]{ 0,  0,  1}{32.17} & 32.11 & 32.16 \\
				Urban100 & 32.13 & \textcolor[rgb]{ 1,  0,  0}{32.35} & 32.28 & 32.24 & 32.01 & \textcolor[rgb]{ 0,  0,  1}{32.29} \\
				Manga109 & 38.47 & \textcolor[rgb]{ 1,  0,  0}{38.86} & \textcolor[rgb]{ 0,  0,  1}{38.72} & 38.60 & 38.20 & 38.62 \\
				Average & 34.40 & \textcolor[rgb]{ 1,  0,  0}{34.63} & \textcolor[rgb]{ 0,  0,  1}{34.53} & 34.48 & 34.25 & 34.50 \\
				\bottomrule
		\end{tabular}}
	\end{table}

\begin{table}[htbp]
	\centering
	\caption{Ablation study inside our ADAM. We report results for image super-resolution with scaling factor $\times$2.}
	\label{tab:ablation inside ADAM}
	\resizebox{6.8cm}{1.1cm}{
		\begin{tabular}{cccccc}
			\toprule
			& Baseline & +ADAM  & +ADAM-C & +ADAM-S & +ADAM-NW \\
			\midrule
			ACAB  & & $\surd$     & $\surd$     &       & $\surd$ \\
			ASAB  & & $\surd$     &       & $\surd$     & $\surd$ \\
			$\alpha$     & & $\surd$     &       & $\surd$     &  \\
			$\beta$     & & $\surd$     & $\surd$     &       &  \\
			PSNR  & 34.40 & 34.63 & 34.48 & 34.51 & 34.54 \\
			\bottomrule
	\end{tabular}}
\end{table}

	\subsubsection{Adaptive hierarchical attention module (AHAM)}
	As shown in Table \ref{tab:ablotionGCFM}, our AHAM obtains a significant improvement for image restoration (the average gain is 0.19dB) with only 1K additional parameters (which is an increase of 0.07\%). Therefore, it can be inserted into most CNN-based image restoration methods.
	\begin{table}[htbp]
		\centering
		\caption{Performance of our AHAM for image super-resolution with scaling factor $\times$2. \textcolor[rgb]{ 1,  0,  0}{Red} color indicates the best performance.}
		\label{tab:ablotionGCFM}
		\resizebox{\linewidth}{0.7cm}{
			\begin{tabular}{cccccccc}
				\toprule
				\multicolumn{1}{c}{} & Params. & \multicolumn{1}{c}{Set5} & \multicolumn{1}{c}{Set14} & \multicolumn{1}{c}{BSD100} & \multicolumn{1}{c}{Urban100} & \multicolumn{1}{c}{Manga109} & \multicolumn{1}{c}{Average} \\
				\midrule
				Baseline & 1370K & 37.90 & 33.58 & 32.17 & 32.13 & 38.47 & 34.40 \\
				\midrule
				+AHAM  & 1371K & \textcolor[rgb]{ 1,  0,  0}{38.11} & \textcolor[rgb]{ 1,  0,  0}{33.69} & \textcolor[rgb]{ 1,  0,  0}{32.22} & \textcolor[rgb]{ 1,  0,  0}{32.30} & \textcolor[rgb]{ 1,  0,  0}{38.81} & \textcolor[rgb]{ 1,  0,  0}{34.59} \\
				\bottomrule
		\end{tabular}}
	\end{table}
	
	\subsection{Image Super-Resolution}
	For image super-resolution, we compare our A-CubeNet with state-of-the-art image super-resolution methods: SRCNN \cite{dong2014learning}, FSRCNN \cite{dong2016accelerating}, VDSR \cite{kim2016accurate}, DRCN \cite{kim2016deeply}, LapSRN \cite{lai2017deep}, DRRN \cite{tai2017image}, MemNet \cite{tai2017memnet}, CARN \cite{ahn2018fast}, FALSR \cite{chu2019fast}, SelNet \cite{choi2017deep}, MoreMNAS \cite{chu2019multi}, SRMDNF \cite{zhang2018learning}, MSRN \cite{li2018multi}, IDN \cite{hui2018fast}, SRRAM \cite{kim2018ram}, IMDN \cite{hui2019lightweight}, SRDenseNet \cite{tong2017image}. All the quantitative results for various scaling factors (e.g., $\times$2, $\times$3, $\times$4) are reported in Table \ref{tab:super-resolution}. As shown in Table \ref{tab:super-resolution}, our A-CubeNet achieves the best PSNR and SSIM results for all scaling factors. \par
	It is worth noting that EDSR \cite{lim2017enhanced}, D-DBPN \cite{haris2018deep}, RDN \cite{zhang2018residual}, RCAN \cite{zhang2018image}, and SAN \cite{dai2019second} have higher performance than our A-CubeNet. However, we do not compare with these models because it is meaningless to compare two models with large differences in parameter and depth. Specifically, the maximum number of parameters of our A-CubeNet is only 1524K, which is much smaller than 43M in EDSR, 16M in RCAN and 15.7M in SAN. The network depth of our A-CubeNet (about 40 convolutional layers) is also much shallower than that of RCAN (about 400 convolutional layers) and SAN (about 400 convolutional layers). \par
\begin{figure}[htbp]
		\centering
		\includegraphics[width=\linewidth]{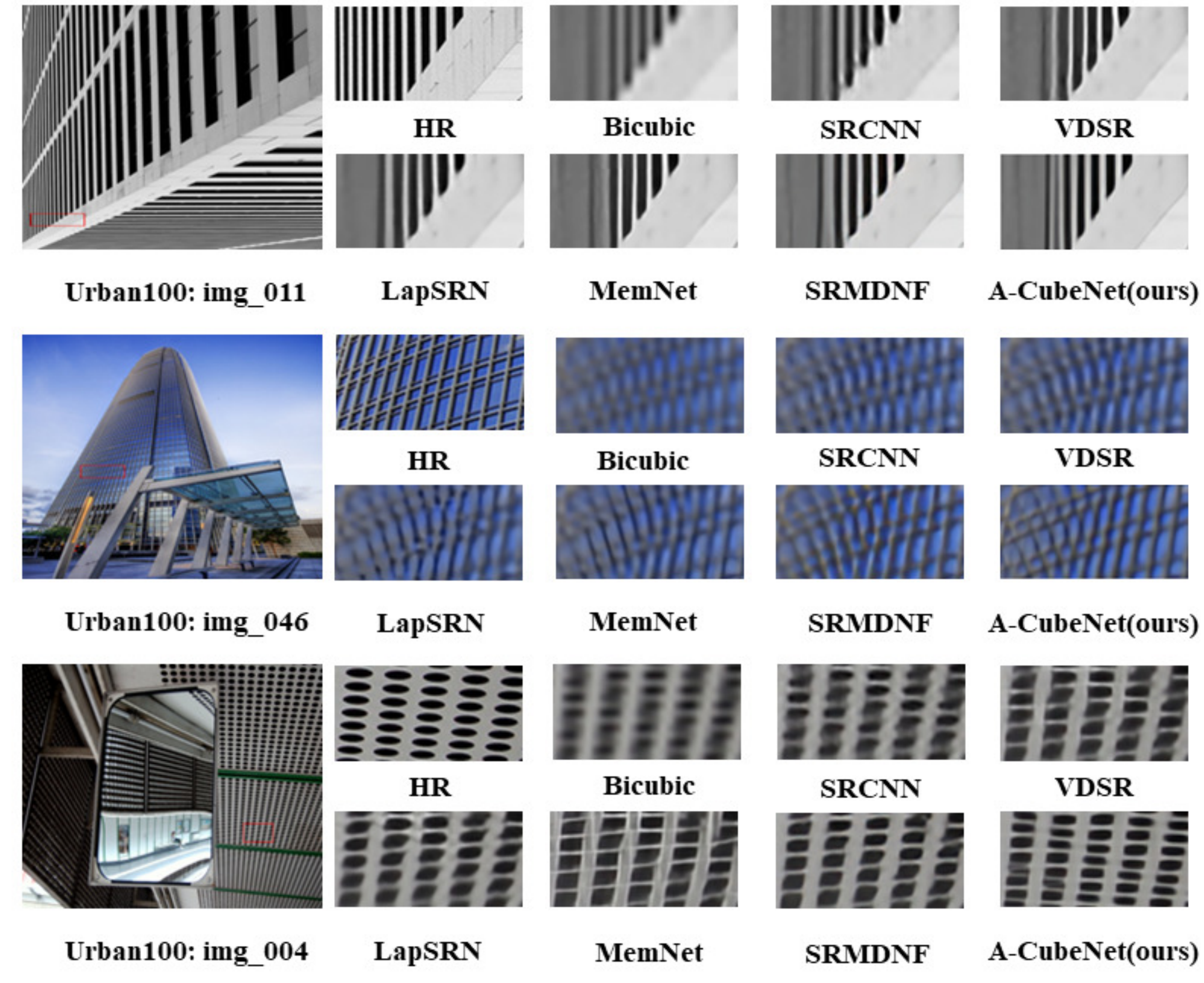}
		\includegraphics[width=\linewidth]{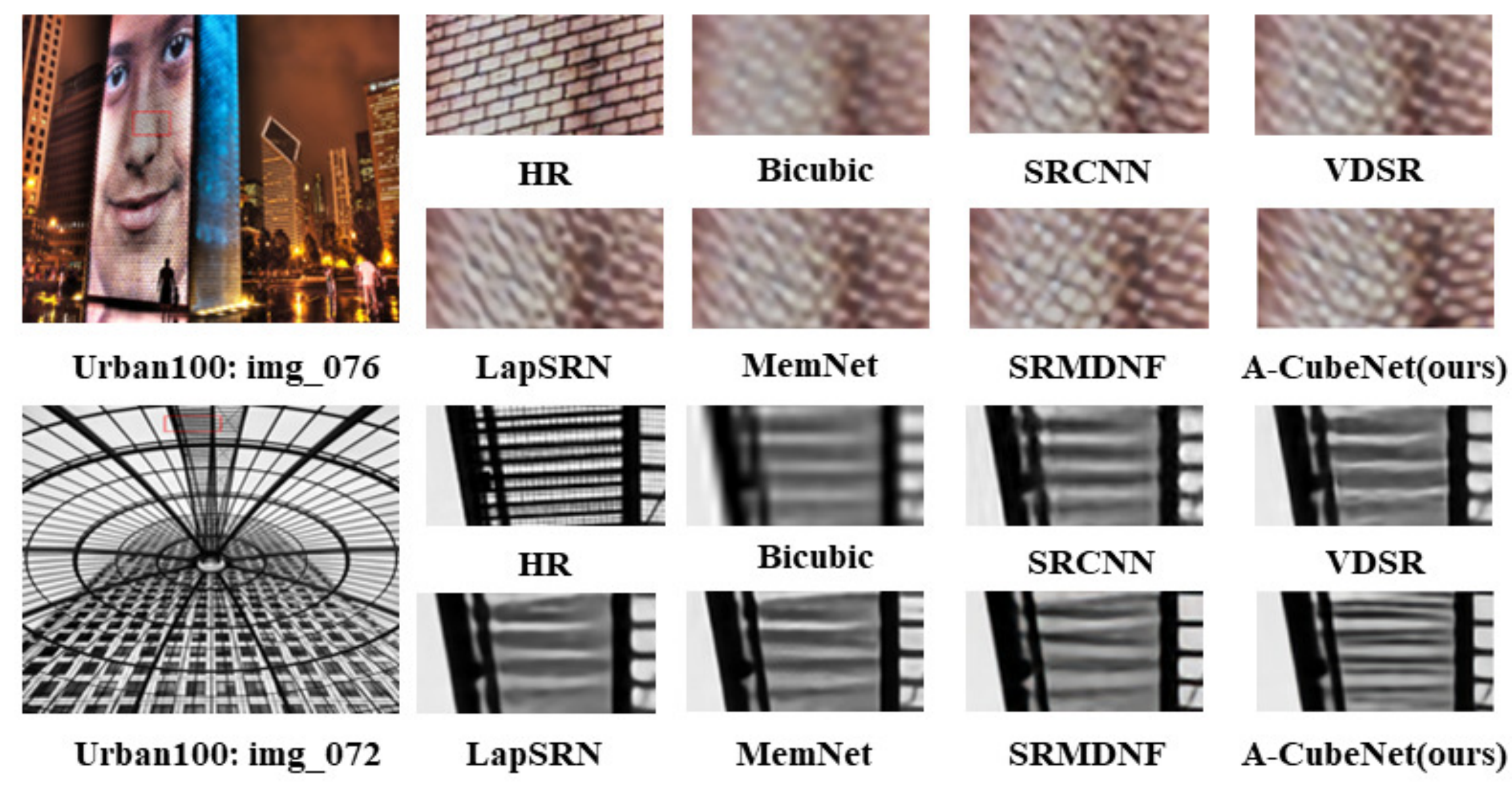}
		\caption{Image super-resolution results with scaling factor $\times$4}
		\Description{Image super-resolution results with scaling factor $\times$4}
		\label{fig:sr}
	\end{figure}

	We further show visual results of different methods in Figure \ref{fig:sr}. Visual results under scaling factor $\times$4 are provided. As shown in Figure \ref{fig:sr}, other methods fail to recover more image details and output heavy blurring artifacts. Compared to these methods, the SR images reconstructed by our A-CubeNet is closer to the HR image in details. These comparisons further demonstrate the effectiveness of our A-CubeNet with the usage of three-dimensional global attention.
	\begin{table}[htbp]
		\centering
		\caption{Quantitative results about image super-resolution. \textcolor[rgb]{ 1,  0,  0}{Red} and \textcolor[rgb]{ 0,  0,  1}{blue} colors indicate the best and second best performance, respectively.}
		\label{tab:super-resolution}
		\resizebox{\linewidth}{7cm}{
			\begin{tabular}{cccccccc}
				\toprule
				\multicolumn{1}{c}{\multirow{2}[4]{*}{Scale}} & \multirow{2}[4]{*}{Model} & \multirow{2}[4]{*}{Params} & Set5  & Set14 & BSD100 & Urban100 & Manga109 \\
				\cmidrule{4-8}          & \multicolumn{1}{c}{} & \multicolumn{1}{c}{} & PSNR/SSIM & PSNR/SSIM & PSNR/SSIM & PSNR/SSIM & PSNR/SSIM \\
				\midrule
				\multirow{21}[2]{*}{2} & Bicubic & -     & \multicolumn{1}{c}{33.66/0.9299} & \multicolumn{1}{c}{30.24/0.8688} & \multicolumn{1}{c}{29.56/0.8431} & \multicolumn{1}{c}{26.88/0.8403} & \multicolumn{1}{c}{30.80/0.9339} \\
				& SRCNN & \multicolumn{1}{c}{57K} & \multicolumn{1}{c}{36.66/0.9542} & \multicolumn{1}{c}{32.42/0.9063} & \multicolumn{1}{c}{31.36/0.8879} & \multicolumn{1}{c}{29.50/0.8946} & \multicolumn{1}{c}{35.74/0.9661} \\
				& FSRCNN & \multicolumn{1}{c}{12K} & \multicolumn{1}{c}{37.00/0.9558} & \multicolumn{1}{c}{32.63/0.9088} & \multicolumn{1}{c}{31.53/0.8920} & \multicolumn{1}{c}{29.88/0.9020} & \multicolumn{1}{c}{36.67/0.9694} \\
				& VDSR  & \multicolumn{1}{c}{665K } & \multicolumn{1}{c}{37.53/0.9587} & \multicolumn{1}{c}{33.03/0.9124} & \multicolumn{1}{c}{31.90/0.8960} & \multicolumn{1}{c}{30.76/0.9140} & \multicolumn{1}{c}{37.22/0.9729} \\
				& DRCN  & \multicolumn{1}{c}{1,774K} & \multicolumn{1}{c}{37.63/0.9588} & \multicolumn{1}{c}{33.04/0.9118} & \multicolumn{1}{c}{31.85/0.8942} & \multicolumn{1}{c}{30.75/0.9133} & \multicolumn{1}{c}{37.63/0.9723} \\
				& LapSRN  & \multicolumn{1}{c}{813K } & \multicolumn{1}{c}{37.52/0.9590} & \multicolumn{1}{c}{33.08/0.9130} & \multicolumn{1}{c}{31.80/0.8950} & \multicolumn{1}{c}{30.41/0.9100} & \multicolumn{1}{c}{37.27/0.9740} \\
				& DRRN  & 297K  & \multicolumn{1}{c}{37.74/0.9591 } & \multicolumn{1}{c}{33.23/0.9136} & \multicolumn{1}{c}{32.05/0.8973} & \multicolumn{1}{c}{31.23/0.9188} & \multicolumn{1}{c}{37.92/0.9760} \\
				& MemNet & \multicolumn{1}{c}{677K} & \multicolumn{1}{c}{37.78/0.9597} & \multicolumn{1}{c}{33.28/0.9142} & \multicolumn{1}{c}{32.08/0.8978} & \multicolumn{1}{c}{31.31/0.9195} & \multicolumn{1}{c}{37.72/0.9740} \\
				& \multicolumn{1}{c}{CARN-M} & \multicolumn{1}{c}{412K} & \multicolumn{1}{c}{37.53/0.9583} & \multicolumn{1}{c}{33.26/0.9141} & \multicolumn{1}{c}{31.92/0.8960 } & \multicolumn{1}{c}{31.23/0.9193} & \multicolumn{1}{c}{-} \\
				& \multicolumn{1}{c}{FALSR-B} & \multicolumn{1}{c}{326k} & \multicolumn{1}{c}{37.61/0.9585} & \multicolumn{1}{c}{33.29/0.9143} & \multicolumn{1}{c}{31.97/0.8967} & \multicolumn{1}{c}{31.28/0.9191} & \multicolumn{1}{c}{-} \\
				& \multicolumn{1}{c}{FALSR-C} & \multicolumn{1}{c}{408k} & \multicolumn{1}{c}{37.66/0.9586 } & \multicolumn{1}{c}{33.26/0.9140} & \multicolumn{1}{c}{31.96/0.8965} & \multicolumn{1}{c}{31.24/0.9187} & \multicolumn{1}{c}{-} \\
				& \multicolumn{1}{c}{SelNet} & \multicolumn{1}{c}{974K} & \multicolumn{1}{c}{37.89/0.9598} & \multicolumn{1}{c}{33.61/0.9160} & \multicolumn{1}{c}{32.08/0.8984} & \multicolumn{1}{c}{-} & \multicolumn{1}{c}{-} \\
				& \multicolumn{1}{c}{MoreMNAS} & \multicolumn{1}{c}{1,039K} & \multicolumn{1}{c}{37.63/0.9584} & \multicolumn{1}{c}{33.23/0.9138} & \multicolumn{1}{c}{31.95/0.8961} & \multicolumn{1}{c}{31.24/0.9187} & \multicolumn{1}{c}{-} \\
				& \multicolumn{1}{c}{ FALSR-A} & \multicolumn{1}{c}{1,021K} & \multicolumn{1}{c}{37.82/0.9595} & \multicolumn{1}{c}{33.55/0.9168 } & \multicolumn{1}{c}{32.12/0.8987 } & \multicolumn{1}{c}{31.93/0.9256} & \multicolumn{1}{c}{-} \\
				& \multicolumn{1}{c}{SRMDNF} & \multicolumn{1}{c}{1,513K} & \multicolumn{1}{c}{37.79/0.9600} & \multicolumn{1}{c}{33.32/0.9150} & \multicolumn{1}{c}{32.05/0.8980} & \multicolumn{1}{c}{31.33/0.9200} & \multicolumn{1}{c}{38.07/0.9761} \\
				& \multicolumn{1}{c}{CARN} & \multicolumn{1}{c}{1,592K} & \multicolumn{1}{c}{37.76/0.9590} & \multicolumn{1}{c}{33.52/0.9166} & \multicolumn{1}{c}{32.09/0.8978} & \multicolumn{1}{c}{31.92/0.9256} & \multicolumn{1}{c}{38.36/0.9765} \\
				& MSRN  & \multicolumn{1}{c}{5,930K} & \multicolumn{1}{c}{\textcolor[rgb]{ 0,  0,  1}{38.08}/\textcolor[rgb]{ 0,  0,  1}{0.9607}} & \multicolumn{1}{c}{\textcolor[rgb]{ 0,  0,  1}{33.70}/\textcolor[rgb]{ 0,  0,  1}{0.9186}} & \multicolumn{1}{c}{\textcolor[rgb]{ 0,  0,  1}{32.23}/\textcolor[rgb]{ 0,  0,  1}{0.9002}} & \multicolumn{1}{c}{\textcolor[rgb]{ 0,  0,  1}{32.29}/\textcolor[rgb]{ 0,  0,  1}{0.9303}} & \multicolumn{1}{c}{\textcolor[rgb]{ 0,  0,  1}{38.69}/0.9772} \\
				& IDN   & \multicolumn{1}{c}{553K} & \multicolumn{1}{c}{37.83/0.9600} & \multicolumn{1}{c}{33.30/0.9148} & \multicolumn{1}{c}{32.08/0.8985} & \multicolumn{1}{c}{31.27/0.9196} & \multicolumn{1}{c}{38.01/0.9749} \\
				& SRRAM & \multicolumn{1}{c}{942K} & \multicolumn{1}{c}{37.82/0.9592} & \multicolumn{1}{c}{33.48/0.9171} & \multicolumn{1}{c}{32.12/0.8983} & \multicolumn{1}{c}{32.05/0.9264} & \multicolumn{1}{c}{-} \\
				& IMDN  & \multicolumn{1}{c}{694K} & \multicolumn{1}{c}{38.00/0.9605} & \multicolumn{1}{c}{33.63/0.9177} & \multicolumn{1}{c}{32.19/0.8996} & \multicolumn{1}{c}{32.17/0.9283} & \multicolumn{1}{c}{\textcolor[rgb]{ 1,  0,  0}{38.88}/\textcolor[rgb]{ 0,  0,  1}{0.9774}} \\
				& A-CubeNet(Ours) & 1376K & \textcolor[rgb]{ 1,  0,  0}{38.12}/\textcolor[rgb]{ 1,  0,  0}{0.9609} & \textcolor[rgb]{ 1,  0,  0}{33.73}/\textcolor[rgb]{ 1,  0,  0}{0.9191} & \textcolor[rgb]{ 1,  0,  0}{32.26}/\textcolor[rgb]{ 1,  0,  0}{0.9007} & \textcolor[rgb]{ 1,  0,  0}{32.39}/\textcolor[rgb]{ 1,  0,  0}{0.9308} & \textcolor[rgb]{ 1,  0,  0}{38.88}/\textcolor[rgb]{ 1,  0,  0}{0.9776} \\
				\midrule
				\multirow{16}[2]{*}{3} & Bicubic & \multicolumn{1}{c}{-} & \multicolumn{1}{c}{30.39/0.8682} & \multicolumn{1}{c}{27.55/0.7742} & \multicolumn{1}{c}{27.21/0.7385} & \multicolumn{1}{c}{24.46/0.7349} & \multicolumn{1}{c}{26.95/0.8556} \\
				& SRCNN & \multicolumn{1}{c}{57K} & \multicolumn{1}{c}{32.75/0.9090} & \multicolumn{1}{c}{29.28/0.8209} & \multicolumn{1}{c}{28.41/0.7863} & \multicolumn{1}{c}{26.24/0.7989} & \multicolumn{1}{c}{30.59/0.9107} \\
				& FSRCNN & \multicolumn{1}{c}{12K} & \multicolumn{1}{c}{33.16/0.9140} & \multicolumn{1}{c}{29.43/0.8242 } & \multicolumn{1}{c}{28.53/0.7910} & \multicolumn{1}{c}{26.43/0.8080} & \multicolumn{1}{c}{30.98/0.9212} \\
				& VDSR  & \multicolumn{1}{c}{665K } & \multicolumn{1}{c}{33.66/0.9213} & \multicolumn{1}{c}{29.77/0.8314} & \multicolumn{1}{c}{28.82/0.7976} & \multicolumn{1}{c}{27.14/0.8279} & \multicolumn{1}{c}{32.01/0.9310} \\
				& DRCN  & \multicolumn{1}{c}{1,774K} & \multicolumn{1}{c}{33.82/0.9226} & \multicolumn{1}{c}{29.76/0.8311} & \multicolumn{1}{c}{28.80/0.7963} & \multicolumn{1}{c}{27.15/0.8276} & \multicolumn{1}{c}{32.31/0.9328} \\
				& DRRN  & \multicolumn{1}{c}{297K} & \multicolumn{1}{c}{34.03/0.9244} & \multicolumn{1}{c}{29.96/0.8349} & \multicolumn{1}{c}{28.95/0.8004} & \multicolumn{1}{c}{27.53/0.8378} & \multicolumn{1}{c}{32.74/0.9390} \\
				& MemNet & \multicolumn{1}{c}{677K} & \multicolumn{1}{c}{34.09/0.9248} & \multicolumn{1}{c}{30.00/0.8350} & \multicolumn{1}{c}{28.96/0.8001} & \multicolumn{1}{c}{27.56/0.8376} & \multicolumn{1}{c}{32.51/0.9369} \\
				& \multicolumn{1}{c}{CARN-M} & \multicolumn{1}{c}{412K} & \multicolumn{1}{c}{33.99/0.9236} & \multicolumn{1}{c}{30.08/0.8367} & \multicolumn{1}{c}{28.91/0.8000} & \multicolumn{1}{c}{27.55/0.8385} & \multicolumn{1}{c}{-} \\
				& \multicolumn{1}{c}{SelNet} & \multicolumn{1}{c}{1,159K} & \multicolumn{1}{c}{34.27/0.9257} & \multicolumn{1}{c}{30.30/0.8399} & \multicolumn{1}{c}{28.97/0.8025} & \multicolumn{1}{c}{-} & \multicolumn{1}{c}{-} \\
				& \multicolumn{1}{c}{SRMDNF} & \multicolumn{1}{c}{1,530K} & \multicolumn{1}{c}{34.12/0.9250} & \multicolumn{1}{c}{30.04/0.8370} & \multicolumn{1}{c}{28.97/0.8030} & \multicolumn{1}{c}{27.57/0.8400} & \multicolumn{1}{c}{33.00/0.9403} \\
				& \multicolumn{1}{c}{CARN} & \multicolumn{1}{c}{1,592K} & \multicolumn{1}{c}{34.29/0.9255} & \multicolumn{1}{c}{30.29/0.8407} & \multicolumn{1}{c}{29.06/0.8034} & \multicolumn{1}{c}{28.06/0.8493} & \multicolumn{1}{c}{33.50/0.9440} \\
				& MSRN  & \multicolumn{1}{c}{6,114K} & \multicolumn{1}{c}{\textcolor[rgb]{ 0,  0,  1}{34.46}/\textcolor[rgb]{ 0,  0,  1}{0.9278}} & \multicolumn{1}{c}{\textcolor[rgb]{ 0,  0,  1}{30.41}/\textcolor[rgb]{ 0,  0,  1}{0.8437}} & \multicolumn{1}{c}{\textcolor[rgb]{ 0,  0,  1}{29.15}/\textcolor[rgb]{ 0,  0,  1}{0.8064}} & \multicolumn{1}{c}{\textcolor[rgb]{ 0,  0,  1}{28.33}/\textcolor[rgb]{ 0,  0,  1}{0.8561}} & \multicolumn{1}{c}{\textcolor[rgb]{ 0,  0,  1}{33.67}/\textcolor[rgb]{ 0,  0,  1}{0.9456}} \\
				& IDN   & \multicolumn{1}{c}{553K} & \multicolumn{1}{c}{34.11/0.9253} & \multicolumn{1}{c}{29.99/0.8354} & \multicolumn{1}{c}{28.95/0.8013} & \multicolumn{1}{c}{27.42/0.8359} & \multicolumn{1}{c}{32.71/0.9381} \\
				& SRRAM & \multicolumn{1}{c}{1,127K} & \multicolumn{1}{c}{34.30/0.9256} & \multicolumn{1}{c}{30.32/0.8417} & \multicolumn{1}{c}{29.07/0.8039} & \multicolumn{1}{c}{28.12/0.8507} & \multicolumn{1}{c}{-} \\
				& IMDN  & \multicolumn{1}{c}{703K} & \multicolumn{1}{c}{34.36/0.9270} & \multicolumn{1}{c}{30.32/0.8417} & \multicolumn{1}{c}{29.09/0.8046} & \multicolumn{1}{c}{28.17/0.8519} & \multicolumn{1}{c}{33.61/0.9445} \\
				& A-CubeNet(Ours) & 1561K & \textcolor[rgb]{ 1,  0,  0}{34.53}/\textcolor[rgb]{ 1,  0,  0}{0.9281} & \textcolor[rgb]{ 1,  0,  0}{30.45}/\textcolor[rgb]{ 1,  0,  0}{0.8441} & \textcolor[rgb]{ 1,  0,  0}{29.17}/\textcolor[rgb]{ 1,  0,  0}{0.8068} & \textcolor[rgb]{ 1,  0,  0}{28.38}/\textcolor[rgb]{ 1,  0,  0}{0.8568} & \textcolor[rgb]{ 1,  0,  0}{33.90}/\textcolor[rgb]{ 1,  0,  0}{0.9466} \\
				\midrule
				\multirow{18}[2]{*}{4} & Bicubic & -     & \multicolumn{1}{c}{28.42/0.8104} & \multicolumn{1}{c}{26.00/0.7027} & \multicolumn{1}{c}{25.96/0.6675} & \multicolumn{1}{c}{23.14/0.6577} & \multicolumn{1}{c}{24.89/0.7866} \\
				& SRCNN & \multicolumn{1}{c}{57K} & \multicolumn{1}{c}{30.48/0.8628} & \multicolumn{1}{c}{27.49/0.7503} & \multicolumn{1}{c}{26.90/0.7101} & \multicolumn{1}{c}{24.52/0.7221} & \multicolumn{1}{c}{27.66/0.8505} \\
				& FSRCNN & \multicolumn{1}{c}{12K} & \multicolumn{1}{c}{30.71/0.8657} & \multicolumn{1}{c}{27.59/0.7535} & \multicolumn{1}{c}{26.98/0.7150} & \multicolumn{1}{c}{24.62/0.7280} & \multicolumn{1}{c}{27.90/0.8517} \\
				& VDSR  & \multicolumn{1}{c}{665K } & \multicolumn{1}{c}{31.35/0.8838} & \multicolumn{1}{c}{28.01/0.7674} & \multicolumn{1}{c}{27.29/0.7251} & \multicolumn{1}{c}{25.18/0.7524} & \multicolumn{1}{c}{28.83/0.8809} \\
				& DRCN  & \multicolumn{1}{c}{1,774K} & \multicolumn{1}{c}{31.53/0.8854} & \multicolumn{1}{c}{28.02/0.7670} & \multicolumn{1}{c}{27.23/0.7233} & \multicolumn{1}{c}{25.14/0.7510} & \multicolumn{1}{c}{28.98/0.8816} \\
				& LapSRN  & \multicolumn{1}{c}{813K } & \multicolumn{1}{c}{31.54/0.8850} & \multicolumn{1}{c}{28.19/0.7720} & \multicolumn{1}{c}{27.32/0.7280} & \multicolumn{1}{c}{25.21/0.7560} & \multicolumn{1}{c}{29.09/0.8845} \\
				& DRRN  & 297K  & \multicolumn{1}{c}{31.68/0.8888} & \multicolumn{1}{c}{28.21/0.7720} & \multicolumn{1}{c}{27.38/0.7284} & \multicolumn{1}{c}{25.44/0.7638} & \multicolumn{1}{c}{29.46/0.8960} \\
				& MemNet & \multicolumn{1}{c}{677K} & \multicolumn{1}{c}{31.74/0.8893} & \multicolumn{1}{c}{28.26/0.7723} & \multicolumn{1}{c}{27.40/0.7281} & \multicolumn{1}{c}{25.50/0.7630} & \multicolumn{1}{c}{29.42/0.8942} \\
				& \multicolumn{1}{c}{CARN-M} & \multicolumn{1}{c}{412K} & \multicolumn{1}{c}{31.92/0.8903} & \multicolumn{1}{c}{28.42/0.7762} & \multicolumn{1}{c}{27.44/0.7304} & \multicolumn{1}{c}{25.62/0.7694} & \multicolumn{1}{c}{-} \\
				& \multicolumn{1}{c}{SelNet} & \multicolumn{1}{c}{1,417K} & \multicolumn{1}{c}{32.00/0.8931} & \multicolumn{1}{c}{28.49/0.7783} & \multicolumn{1}{c}{27.44/0.7325} & \multicolumn{1}{c}{-} & \multicolumn{1}{c}{-} \\
				& \multicolumn{1}{c}{SRDenseNet} & \multicolumn{1}{c}{2,015K} & \multicolumn{1}{c}{32.02/0.8934} & \multicolumn{1}{c}{28.50/0.7782} & \multicolumn{1}{c}{27.53/0.7337} & \multicolumn{1}{c}{26.05/0.7819} & \multicolumn{1}{c}{-} \\
				& \multicolumn{1}{c}{SRMDNF} & \multicolumn{1}{c}{1,555K} & \multicolumn{1}{c}{31.96/0.8930} & \multicolumn{1}{c}{28.35/0.7770} & \multicolumn{1}{c}{27.49/0.7340 } & \multicolumn{1}{c}{25.68/0.7730} & \multicolumn{1}{c}{30.09/0.9024} \\
				& \multicolumn{1}{c}{CARN} & \multicolumn{1}{c}{1,592K} & \multicolumn{1}{c}{32.13/0.8937} & \multicolumn{1}{c}{28.60/0.7806} & \multicolumn{1}{c}{27.58/0.7349} & \multicolumn{1}{c}{26.07/0.7837} & \multicolumn{1}{c}{30.47/0.9084} \\
				& MSRN  & \multicolumn{1}{c}{6,078K} & \multicolumn{1}{c}{\textcolor[rgb]{ 0,  0,  1}{32.26}/\textcolor[rgb]{ 0,  0,  1}{0.8960}} & \multicolumn{1}{c}{\textcolor[rgb]{ 0,  0,  1}{28.63}/\textcolor[rgb]{ 0,  0,  1}{0.7836}} & \multicolumn{1}{c}{\textcolor[rgb]{ 0,  0,  1}{27.61}/\textcolor[rgb]{ 0,  0,  1}{0.7380}} & \multicolumn{1}{c}{\textcolor[rgb]{ 0,  0,  1}{26.22}/\textcolor[rgb]{ 0,  0,  1}{0.7911}} & \multicolumn{1}{c}{\textcolor[rgb]{ 0,  0,  1}{30.57}/\textcolor[rgb]{ 0,  0,  1}{0.9103}} \\
				& IDN   & \multicolumn{1}{c}{553K} & \multicolumn{1}{c}{31.82/0.8903} & \multicolumn{1}{c}{28.25/0.7730} & \multicolumn{1}{c}{27.41/0.7297} & \multicolumn{1}{c}{25.41/0.7632} & \multicolumn{1}{c}{29.41/0.8942} \\
				& SRRAM & \multicolumn{1}{c}{1,090K} & \multicolumn{1}{c}{32.13/0.8932} & \multicolumn{1}{c}{28.54/0.7800} & \multicolumn{1}{c}{27.56/0.7350} & \multicolumn{1}{c}{26.05/0.7834} & \multicolumn{1}{c}{-} \\
				& IMDN  & \multicolumn{1}{c}{715K} & \multicolumn{1}{c}{32.21/0.8948} & \multicolumn{1}{c}{28.58/0.7811} & \multicolumn{1}{c}{27.56/0.7353} & \multicolumn{1}{c}{26.04/0.7838} & \multicolumn{1}{c}{30.45/0.9075} \\
				& A-CubeNet(Ours) & 1524K & \textcolor[rgb]{ 1,  0,  0}{32.32}/\textcolor[rgb]{ 1,  0,  0}{0.8969} & \textcolor[rgb]{ 1,  0,  0}{28.72}/\textcolor[rgb]{ 1,  0,  0}{0.7847} & \textcolor[rgb]{ 1,  0,  0}{27.65}/\textcolor[rgb]{ 1,  0,  0}{0.7382} & \textcolor[rgb]{ 1,  0,  0}{26.27}/\textcolor[rgb]{ 1,  0,  0}{0.7913} & \textcolor[rgb]{ 1,  0,  0}{30.81}/\textcolor[rgb]{ 1,  0,  0}{0.9114} \\
				\bottomrule
		\end{tabular}}
	\end{table}
	\subsection{Gray Image Denoising}
	For gray image denosing, we generate the degraded images by adding AWGN noise of different levels (e.g., 10, 30, 50, and 70) to clean images. We compare our A-CubeNet with state-of-the-art gray image denoising methods: BM3D \cite{dabov2007image}, TNRD \cite{chen2016trainable}, DnCNN \cite{zhang2017beyond}, MemNet \cite{tai2017memnet}, IRCNN \cite{zhang2017learning}, and FFDNet \cite{zhang2018ffdnet}. As shown in Table \ref{tab:denoise}, our A-CubeNet achieves the best PSNR results with all noise levels.\par
\begin{figure}[htbp]
		\centering
		\includegraphics[width=\linewidth]{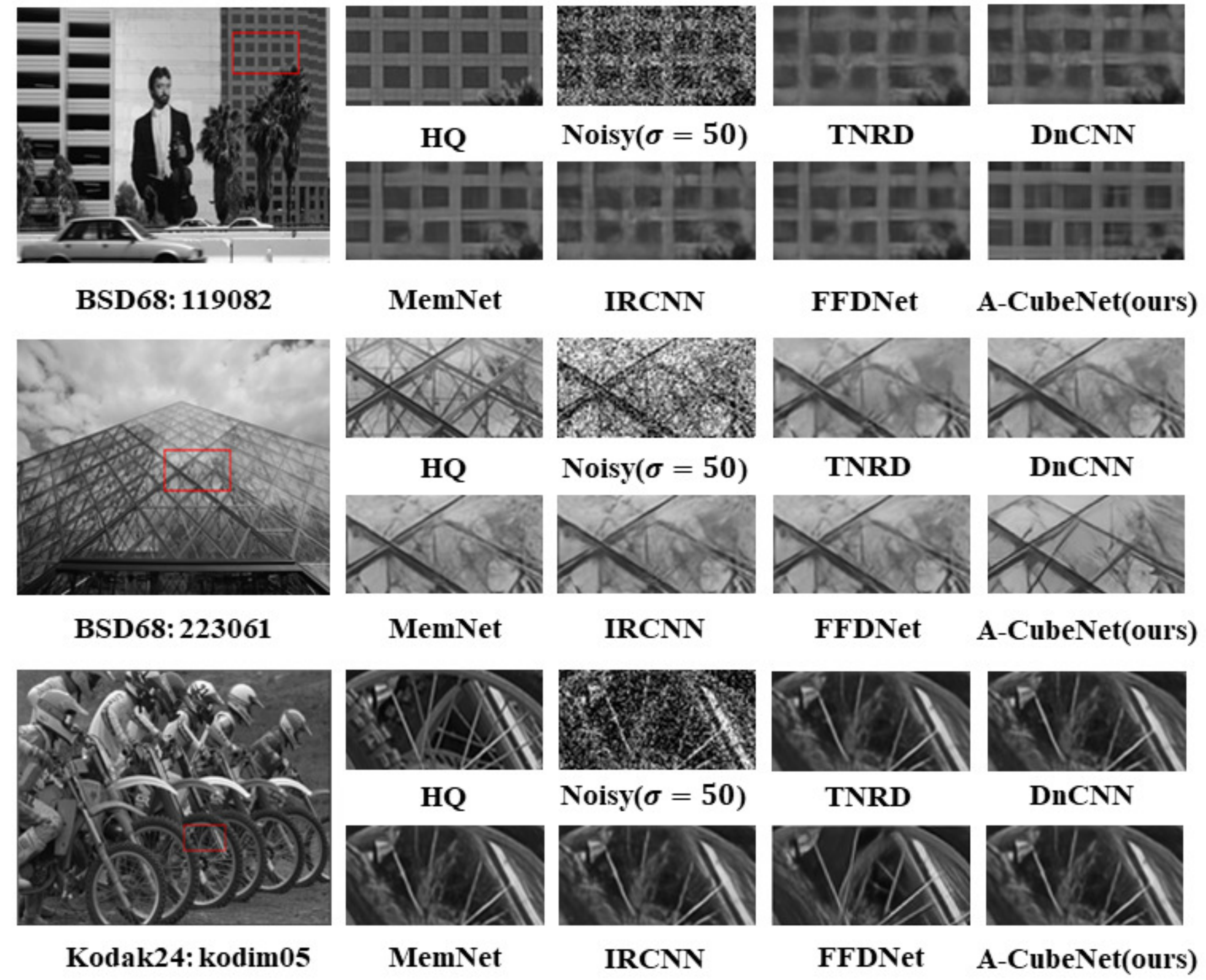}
		\includegraphics[width=\linewidth]{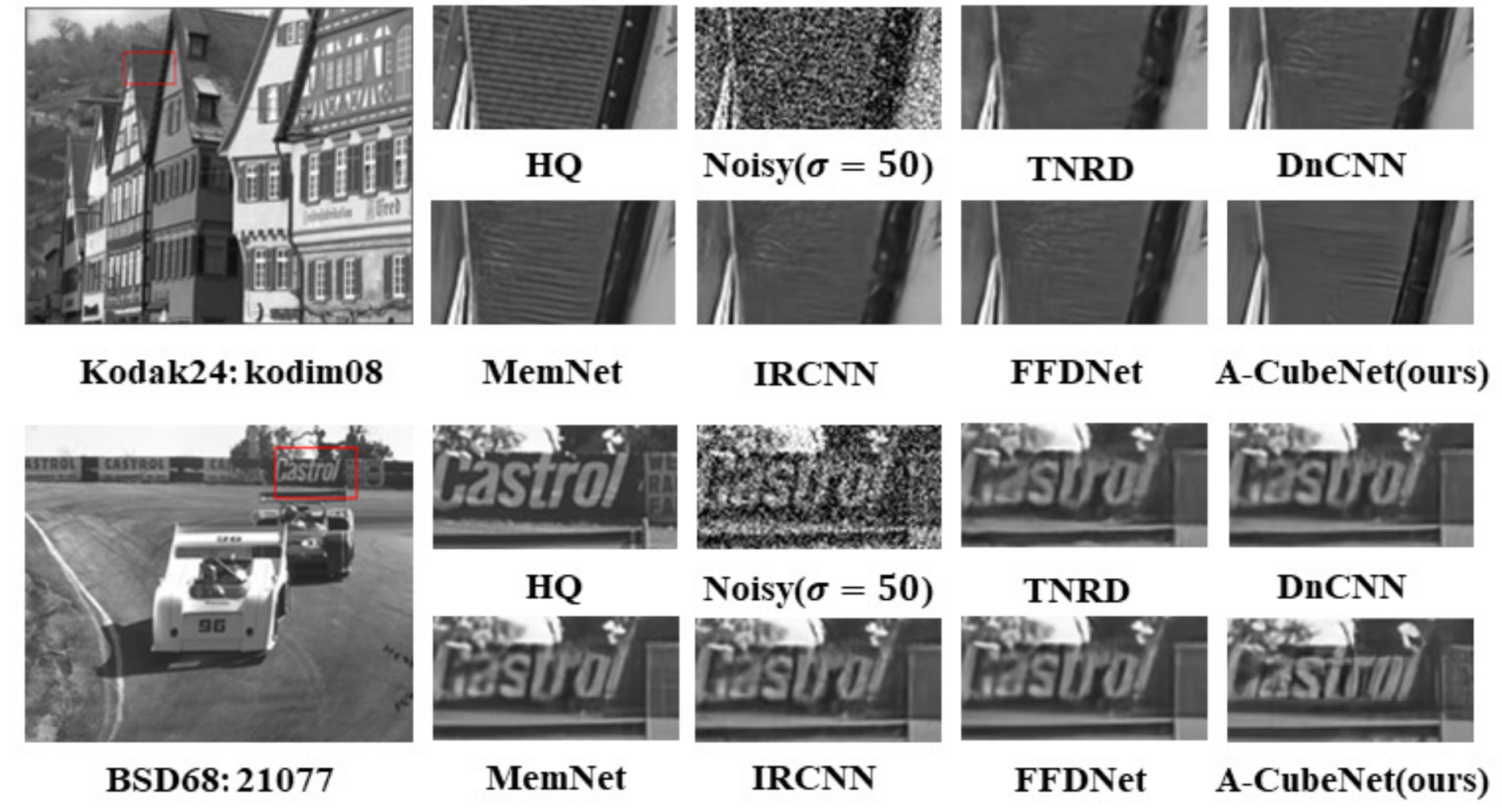}
		\caption{Gray image denoising results with noise level 50}
		\Description{Gray image denoising results with noise level 50}
		\label{fig:dn}
	\end{figure}

	Visual results of different  methods  under noise level $\sigma = 50$ are shown in Figure \ref{fig:dn}. As shown in Figure \ref{fig:dn}, BM3D, TNRD, DnCNN, MemNet, IRCNN, and FFDNet can remove noise to some degree, but also over-smooth some details obviously. Compared to these methods, our A-CubeNet not only removes noise well, but also alleviate over-smoothing artifacts obviously because our A-CubeNet covers the information from the whole image and treats different types of information differently.
	\begin{table}[htbp]
		\centering
		\caption{Quantitative results about gray image denoising. \textcolor[rgb]{ 1,  0,  0}{Red} and \textcolor[rgb]{ 0,  0,  1}{blue} colors indicate the best and second best performance, respectively.}
		\label{tab:denoise}
		\resizebox{\linewidth}{1.55cm}{
			\begin{tabular}{ccccccccc}
				\toprule
				\multirow{2}[4]{*}{Method} & \multicolumn{4}{c}{Kodak24} & \multicolumn{4}{c}{BSD68} \\
				\cmidrule{2-9}    \multicolumn{1}{c}{} & 10    & 30    & 50    & 70    & 10    & 30    & 50    & 70 \\
				\midrule
				BM3D  & 34.39 & 29.13 & 26.99 & 25.73 & 33.31 & 27.76 & 25.62 & 24.44 \\
				TNRD  & 34.41 & 28.87 & 27.20  & 24.95 & 33.41 & 27.66 & 25.97 & 23.83 \\
				DnCNN & \textcolor[rgb]{ 0,  0,  1}{34.90} & 29.62 & 27.51 & 26.08 & \textcolor[rgb]{ 0,  0,  1}{33.88} & 28.36 & 26.23 & 24.90 \\
				MemNet & - & \textcolor[rgb]{ 0,  0,  1}{29.72} & \textcolor[rgb]{ 0,  0,  1}{27.68} & \textcolor[rgb]{ 0,  0,  1}{26.42} & - & \textcolor[rgb]{ 0,  0,  1}{28.43} & \textcolor[rgb]{ 0,  0,  1}{26.35} & \textcolor[rgb]{ 0,  0,  1}{25.09} \\
				IRCNN & 34.76 & 29.53 & 27.45 & - & 33.74 & 28.26 & 26.15 & - \\
				FFDNet & 34.81 & 29.70  & 27.63 & 26.34 & 33.76 & 28.39 & 26.29 & 25.04 \\
				A-CubeNet(Ours) & \textcolor[rgb]{ 1,  0,  0}{35.06} & \textcolor[rgb]{ 1,  0,  0}{29.84} & \textcolor[rgb]{ 1,  0,  0}{27.77} & \textcolor[rgb]{ 1,  0,  0}{26.44} & \textcolor[rgb]{ 1,  0,  0}{33.94} & \textcolor[rgb]{ 1,  0,  0}{28.50} & \textcolor[rgb]{ 1,  0,  0}{26.37} & \textcolor[rgb]{ 1,  0,  0}{25.10} \\
				\bottomrule
		\end{tabular}}
	\end{table}
	\subsection{JPEG Image Deblocking}
	We also apply our A-CubeNet to reduce image compression artifacts. We use the MATLAB JPEG encoder to generate JPEG deblocking inputs with four JPEG quality settings q = 10, 20, 30, 40. For a fair comparison, we perform training and evaluating both on Y channel of the YCbCr color space. We compare our A-CubeNet with SA-DCT \cite{Foi2007Pointwise}, ARCNN \cite{dong2015compression}, TNRD \cite{chen2016trainable}, and DnCNN \cite{zhang2017beyond}. As shown in Table \ref{tab:deblock}, our A-CubeNet achieves the best PSNR and SSIM results with all JPEG quality settings. \par
\begin{figure}[htbp]
		\centering
		\includegraphics[width=\linewidth]{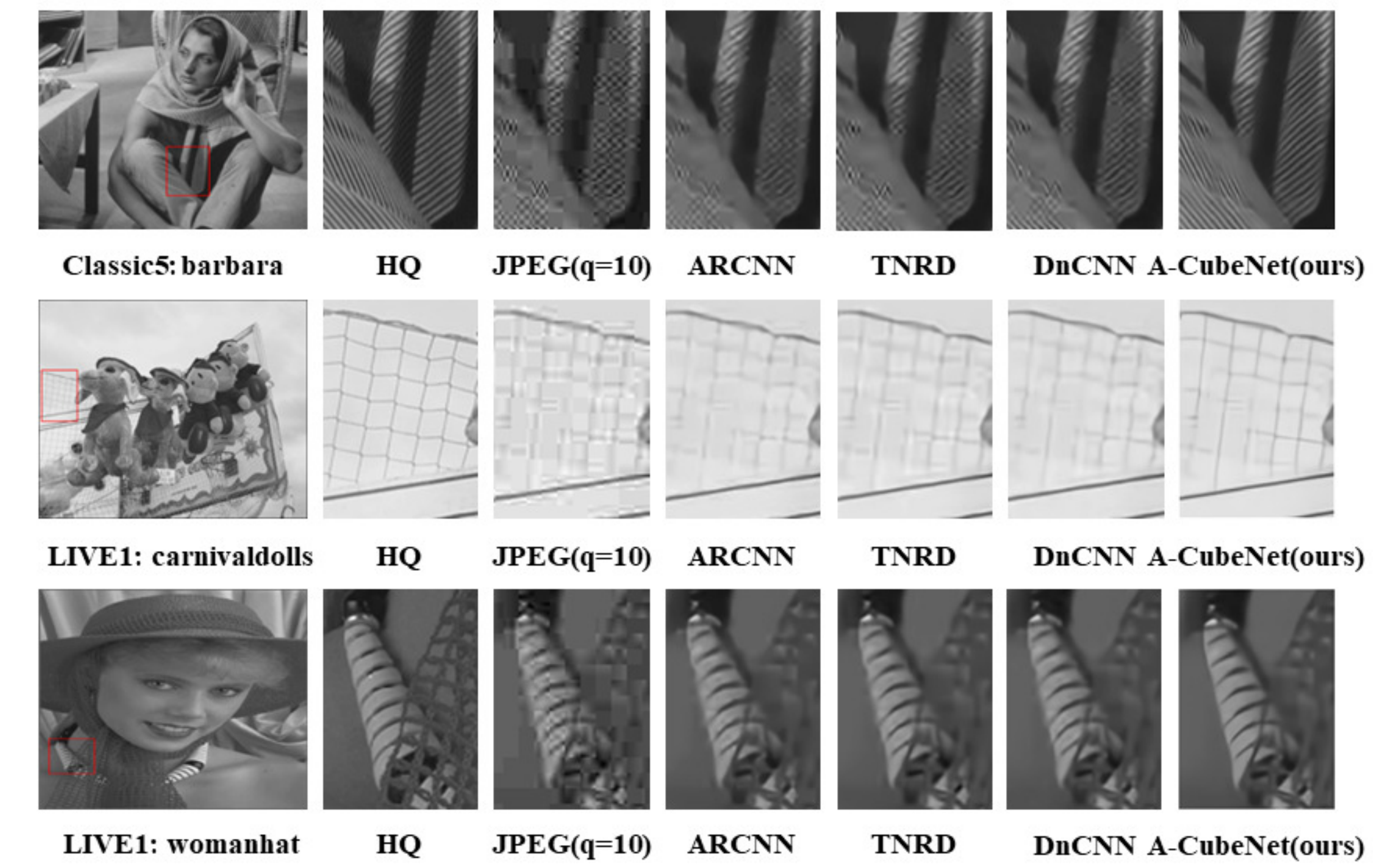}
		\includegraphics[width=\linewidth]{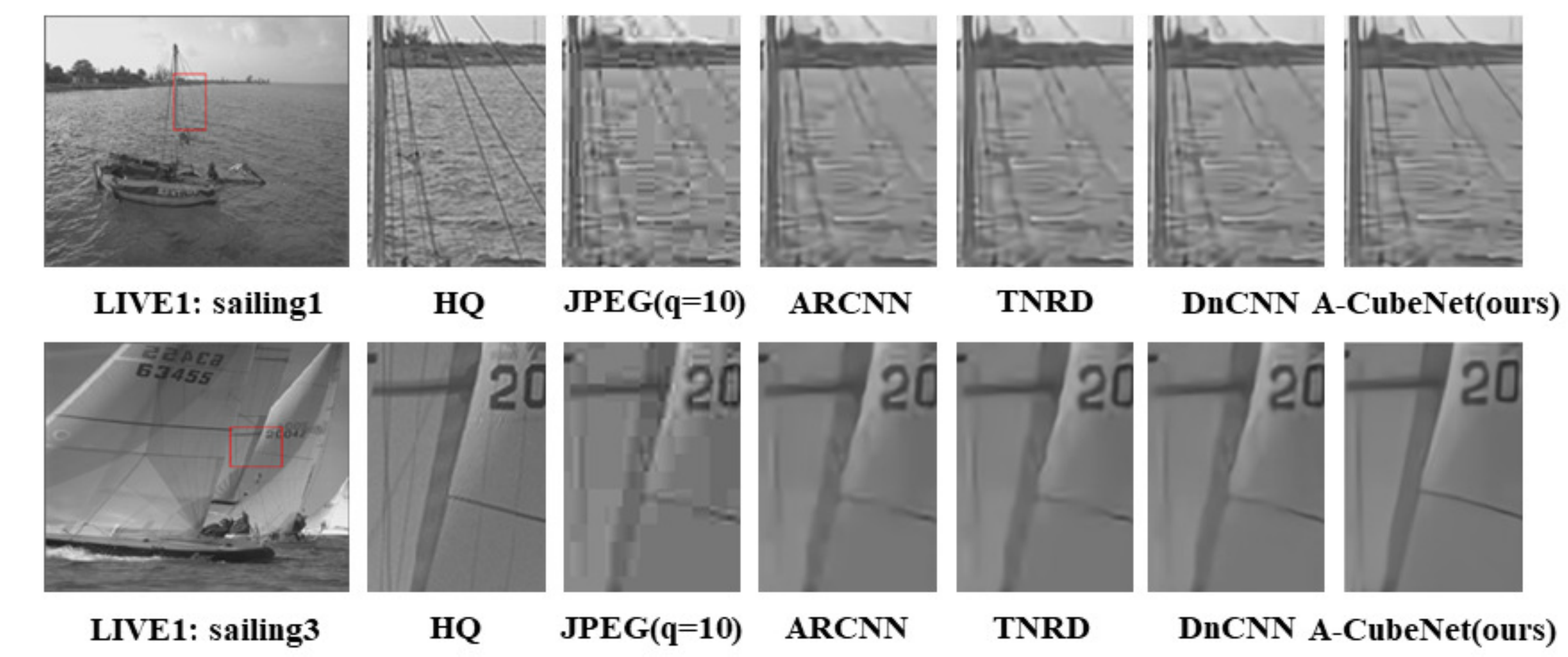}
		\caption{JPEG image deblocking results with JPEG quality 10}
		\Description{JPEG image deblocking results with JPEG quality 10}
		\label{fig:db}
	\end{figure}

	In Figure \ref{fig:db}, we also show visual results of different methods. Visual results under very low JPEG quality (q = 10) are provided. As shown in Figure \ref{fig:db}, SA-DCT, ARCNN, TNRD, and DnCNN can remove artifacts to some degree, but these methods also over-smooth some details. Compared with these methods, our A-CubeNet not only removes artifacts well, but also preserves more details because our A-CubeNet obtains more details with consistent structures by considering three-dimensional global attention.
	\begin{table}[htbp]
		\centering
		\caption{Quantitative results about JPEG image deblocking. \textcolor[rgb]{ 1,  0,  0}{Red} and \textcolor[rgb]{ 0,  0,  1}{blue} colors indicate the best and second best performance, respectively.}
		\label{tab:deblock}
		\resizebox{\linewidth}{1.4cm}{
			\begin{tabular}{cccccccc}
				\toprule
				\multicolumn{1}{c}{\multirow{2}[4]{*}{Dataset}} & \multicolumn{1}{c}{\multirow{2}[4]{*}{q}} & JPEG  & SA-DCT & ARCNN & TNRD  & DnCNN & A-CubeNet(Ours) \\
				\cmidrule{3-8}          &       & PSNR/SSIM & PSNR/SSIM & PSNR/SSIM & PSNR/SSIM & PSNR/SSIM & PSNR/SSIM \\
				\midrule
				\multicolumn{1}{c}{\multirow{4}[2]{*}{LIVE1}} & 10    & 27.77/0.7905 & 28.65/0.8093 & 28.96/0.8076 & 29.15/0.8111 & \textcolor[rgb]{ 0,  0,  1}{29.19}/\textcolor[rgb]{ 0,  0,  1}{0.8123} & \textcolor[rgb]{ 1,  0,  0}{29.54}/\textcolor[rgb]{ 1,  0,  0}{0.8216} \\
				& 20    & 30.07/0.8683 & 30.81/0.8781 & 31.29/0.8733 & 31.46/0.8769 & \textcolor[rgb]{ 0,  0,  1}{31.59}/\textcolor[rgb]{ 0,  0,  1}{0.8802} & \textcolor[rgb]{ 1,  0,  0}{31.93}/\textcolor[rgb]{ 1,  0,  0}{0.8859} \\
				& 30    & 31.41/0.9000 & 32.08/0.9078 & 32.67/0.9043 & 32.84/0.9059 & \textcolor[rgb]{ 0,  0,  1}{32.98}/\textcolor[rgb]{ 0,  0,  1}{0.9090} & \textcolor[rgb]{ 1,  0,  0}{33.35}/\textcolor[rgb]{ 1,  0,  0}{0.9136} \\
				& 40    & 32.35/0.9173 & 32.99/0.9240 & 33.63/0.9198 & -/-   & \textcolor[rgb]{ 0,  0,  1}{33.96}/\textcolor[rgb]{ 0,  0,  1}{0.9247} & \textcolor[rgb]{ 1,  0,  0}{34.36}/\textcolor[rgb]{ 1,  0,  0}{0.9289} \\
				\midrule
				\multicolumn{1}{c}{\multirow{4}[2]{*}{Classic5}} & 10    & 27.82/0.7800 & 28.88/0.8071 & 29.03/0.7929 & 29.28/0.7992 & \textcolor[rgb]{ 0,  0,  1}{29.40}/\textcolor[rgb]{ 0,  0,  1}{0.8026} & \textcolor[rgb]{ 1,  0,  0}{29.84}/\textcolor[rgb]{ 1,  0,  0}{0.8147} \\
				& 20    & 30.12/0.8541 & 30.92/0.8663 & 31.15/0.8517 & 31.47/0.8576 & \textcolor[rgb]{ 0,  0,  1}{31.63}/\textcolor[rgb]{ 0,  0,  1}{0.8610} & \textcolor[rgb]{ 1,  0,  0}{32.04}/\textcolor[rgb]{ 1,  0,  0}{0.8677} \\
				& 30    & 31.48/0.8844 & 32.14/0.8914 & 32.51/0.8806 & 32.78/0.8837 & \textcolor[rgb]{ 0,  0,  1}{32.91}/\textcolor[rgb]{ 0,  0,  1}{0.8861} & \textcolor[rgb]{ 1,  0,  0}{33.30}/\textcolor[rgb]{ 1,  0,  0}{0.8911} \\
				& 40    & 32.43/0.9011 & 33.00/0.9055 & 33.34/0.8953 & -/-   & \textcolor[rgb]{ 0,  0,  1}{33.77}/\textcolor[rgb]{ 0,  0,  1}{0.9003} & \textcolor[rgb]{ 1,  0,  0}{34.16}/\textcolor[rgb]{ 1,  0,  0}{0.9048} \\
				\bottomrule
		\end{tabular}}
	\end{table}
	\section{Conclusion}
	In this paper, we propose a novel attention cube network based on the adaptive dual attention module (ADAM) and  the adaptive hierarchical attention module (AHAM) for high-quality image restoration. These two modules constitute the attention cube from spatial, channel-wise and hierarchical dimensions. Our ADAM can capture the long-range contextual information between pixels and channels. As a result, this module successfully expands the receptive field and effectively distinguishes different types of information. Additionally, our AHAM can capture the long-range hierarchical contextual information to combine different feature maps by weights depending on global context. The ADAM and AHAM cooperate to form an "attention in attention" structure, which is very helpful for image restoration. Our method achieves state-of-the-art image restoration results. In the future, this method will be extended to other image restoration tasks.
	\begin{acks}
	This work was supported by the Natural Science Foundation of Guangdong Province (No. 2020A1515010711), the Natural Science Foundation of China (Nos. 61771276) and the Special Foundation for the Development of Strategic Emerging Industries of Shenzhen (No. JCYJ20170817161845824).
	\end{acks}
\newpage
	\bibliographystyle{ACM-Reference-Format}
	\bibliography{ACM-cit}
\end{document}